\documentclass[useAMS,usenatbib]{articulo}
\usepackage{graphics,epsfig}
\normalsize

\bibpunct[,]{(}{)}{;}{a}{,}{,}

\begin{document}

\title[Transport across an island wake]
{Kinematic studies of transport across an island wake, with
application to the Canary islands.}

\author[Sandulescu et al.]{by
MATHIAS SANDULESCU$^1$, EMILIO HERN{\'A}NDEZ-GARC{\'I}A$^2$,
CRIST\'OBAL L{\'O}PEZ$^2$\thanks{Corresponding
author.\hfil\break e-mail: clopez@imedea.uib.es}, ULRIKE FEUDEL$^1$
\affiliation{
$^{1}$Carl-von-Ossietzky Universit\"at Oldenburg
D-26111 Oldenburg, Germany;
$^{2}$Instituto Mediterr{\'a}neo de Estudios Avanzados,
IMEDEA (CSIC - Universitat de les Illes Balears)
E-07122 Palma de Mallorca, Spain}
}


\maketitle

\begin{abstract}

Transport from nutrient-rich coastal upwellings is a key factor
influencing biological activity in surrounding waters and even in
the open ocean. The rich upwelling in the North-Western African
coast is known to interact strongly with the wake of the Canary
islands, giving rise to filaments and other mesoscale structures
of increased productivity. Motivated by this scenario, we
introduce a simplified two-dimensional kinematic flow describing
the wake of an island in a stream, and study the conditions under
which there is a net transport of substances across the wake. For
small vorticity values in the wake, it acts as a barrier, but
there is a transition when increasing vorticity so that for values
appropriate to the Canary area, it entrains fluid and enhances
cross-wake transport.

\end{abstract}
\section{Introduction}

Chaotic transport in hydrodynamic flows
\cite{Aref2002,Ottino1989,Wiggins1992} is a subject generating a
great amount of interest both in its fundamental aspects and in
its applications to industrial, laboratory, and environmental
flows. A class of problems of particular relevance in the context
of ocean modelling is the one of transport across jets
\citep{Bower1991,Samelson1992,Meyers1994,Rogerson1999,Cencini1999}.
One of the outcomes of these studies is that ocean jets can
behave, depending on parameter regimes, both as barriers to the
transport of the particles and as mixing enhancers, increasing the
interchange of water masses across them \cite{Bower1985}.
Typically there is an increased fluid transport when enhancing the
time dependence of the jet, associated to an increased chaotic
behavior of the fluid trajectories.

\begin{figure}
\mbox{
\includegraphics [width=\textwidth]{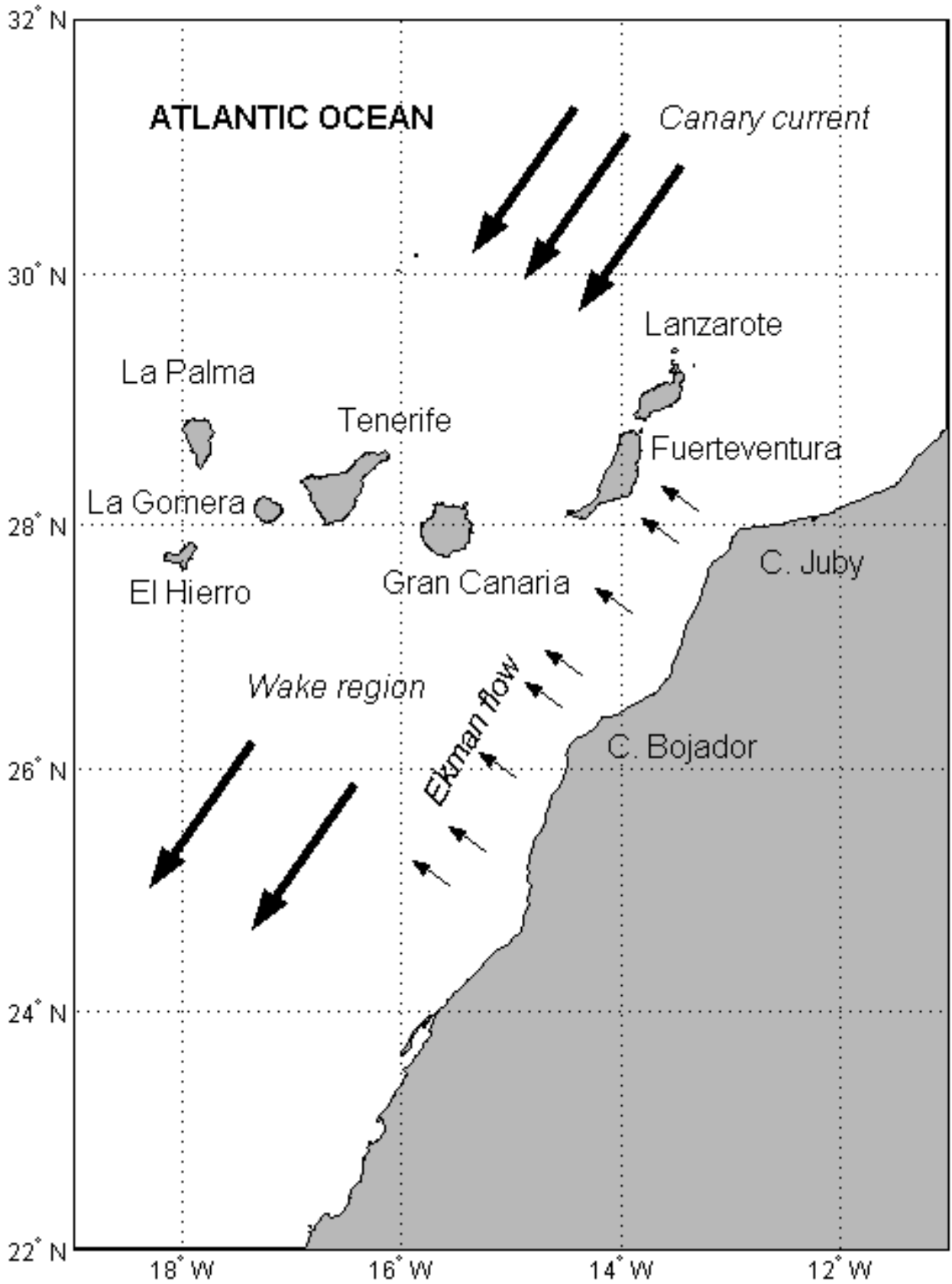}
} \caption{The Canary Islands region, with the Canary current
running southwestwards parallel to the African coast, where there
is an intense upwelling, and impinging on the islands.}
\label{fig:Canary}

\end{figure}

In this Paper we consider a related issue, namely that of fluid
transport across a wake. The motivation arises from situations
occurring in front of the Canary upwelling zone in the Northwest
African coast
\cite{Aristegui1997,Barton1998,Barton2004,Pelegri2005} (See Fig.
\ref{fig:Canary}). There is a strong mesoscale activity in the
wake of the Canary Islands, originated from the current impinging
on them from the North, and running southwards or southwestwards
more or less parallel to the African coast. Wind stress in the lee
region of the islands plays also a role in the generation of
mesoscale eddies \cite{Aristegui1997,Barton1998}. At the same time
there is intense upwelling of depth nutrient-rich waters at the
African coast which is produced by winds parallel to the coast via
the Ekman mechanism. These two systems interact giving rise to
filaments of great biological productivity and complex dynamics.
Our aim in this Paper is to explore a very simple kinematic
mechanism for the formation of such filaments: entrainment by the
wake. We will also determine whether the wake will act as a
barrier, i.e. it will stop the flux of nutrient-rich water towards
the ocean interior, or rather cross-wake transport will be
increased by the presence of eddies. In the second case, which is
the one realized in our model for parameter values appropriate for
the Canary zone, the mechanism may be important to enhance
biological productivity of ocean regions relatively far from the
coastal upwelling.  In any case, we stress that in this work
we focus on the transport of particles into and across the wake
and not on the long-range transport that could drive them far
apart from it.

In general, one can identify three possible mechanisms
contributing to the horizontal transport from a coastal upwelling
across a wake: The first one is the direct effect of Ekman pumping
that transports the upwelled waters in the direction opposite to
the coast. In the second one, coastal water parcels become
entrained by the wake, which stretches and deforms them into
filamental features until some parts reach the ocean interior.
Third, coastal waters may become captured inside eddies, which can
travel long distances.

All of the three mechanisms require the wake to be permeable to
fluid trajectories, and chaotic advection behind the island to be
strong enough to allow transverse transport across the main
current. There have been studies of chaotic transport in flows
modelling island wakes \cite{Miller2002}, but the emphasis was not
in transverse transport. Transport of coastal waters inside eddies
and filaments has been observed in the Canary area
\cite{Aristegui1997,Barton1998,Barton2004,Pelegri2005}.
Nevertheless few attention has been devoted to the relative
importance and interplay between the first two mechanisms. With
numerical solutions of the Navier-Stokes equations in two
dimensions \cite{Duan1997,Shariff1992}, it has been shown, for the
wake behind a cylinder, that an important increase of cross-wake
transport occurs in the Reynolds number range $100-200$. The
phenomenon has been studied in detail \cite{Duan1997}, and has
been associated to topological changes in the structure of the
wake, which allows lobes of fluid to be stretched into filaments
that cross the wake. The similarity of this mechanism to what it
is seen in the Canary area motivates our study, in which we try to
identify an analogous mechanism in a geophysical setting. In
particular, we will show a transition from a situation with a
barrier that does not allow particles to cross the wake, and
another one without barrier, where there is a net transport of
matter across it.  In the real ocean, phenomena such as eddy
detachment and additional filamentation produced by hyperbolic
regions in the neighborhood, can collaborate with the
wake-crossing mechanism reported in this work to produce
long-range transport. But these effects are absent in our model,
and as already mentioned, we focus on the possibility of crossing
the wake, i.e., in the fact that the particles visit the side of
the wake opposite to the place from which they are released. 

With this aim we use in this Paper a kinematic approach to analyze
the interplay between the mechanisms of Ekman transport and
entraining by the wake. We focus on horizontal transport on upper
ocean layers by using a two-dimensional flow, and set up a model
streamfunction having the qualitative features of the wake behind
an island by modifying the streamfunction introduced in
\cite{Jung1993,Ziemniak1994} to model the wake behind a cylinder.
Parameters are chosen in such a way that the relevant geometric
features (sizes, time scales, speeds, ...) are comparable with the
real situation in the Canary Islands zone. We do not expect this
to be an accurate model of the real ocean dynamics, but since the
spatial and temporal scales are taken from observations, we expect
it to capture the correct kinematics of the transport and the
relative importance of the mechanisms involved.

In the following, we first discuss the properties of the velocity
field used, and then characterize the amount of transport in
several parameter regimes. To quantify it we define in the system
an area outside the wake providing a continuous source of
particles, and count how many of them are able to cross the wake
for different parameter values. In our interpretation of the model
as a representation of the Canary zone, the particle source area
is intended to represent the upwelling water close to the African
coast.

The Paper is organized as follows. In the next section we present
the kinematic flow differentiating the three situations that we
want to study: periodic flow, non-periodic flow and periodic flow
with turbulent diffusion of the particles. Then in Section 4 we
briefly comment on the dynamics of the particles. Section 5
contains the results of our work and in Section 6 we write down
our conclusions.

\section{An analytical model for the flow in the wake of an island}
\label{modelo}
Full hydrodynamic simulations of flows in two or three spatial
dimensions involve solving Navier-Stokes equations or
approximations to it. In geophysical contexts, simplified
turbulence closing schemes should be used to simulate the small
unresolved scales. A simple alternative from which considerable
insight has been gained in the past
\citep{Bower1991,Samelson1992,Meyers1994,Rogerson1999,Cencini1999}
is to consider, in two-dimensional incompressible situations, a
model streamfunction $\Psi(x,y,t)$ giving a flow qualitatively
similar to the one under study. The velocity components in $x$-
and $y$-direction and the equations of motion of fluid elements
are:
\begin{eqnarray}
 \dot x &=& v_{x}(x,y,t) = \frac{\partial}{\partial y} \Psi(x,y,t),\nonumber \\
 \dot y &=& v_{y}(x,y,t) = -\frac{\partial}{\partial x} \Psi(x,y,t).
\label{Strfcomp}
\end{eqnarray}

We are interested in the transport perpendicular to the vortex
street in the wake of an island. To keep the geometry of the
island as simple as possible, we assume it to have a circular
shape. Of course, this is a crude approximation to the Canary
islands archipelago. However, observations report on the existence
of vortex streets in the south of the islands which qualitatively
can be understood as emerging from a single large obstacle. As in
\cite{Aristegui1997}, the island of Gran Canaria will be chosen
here as the representative obstacle of the whole archipelago. Our
streamfunction is based on the one introduced in \cite{Jung1993}
and \cite{Ziemniak1994}, but we add to it the effect of Ekman
pumping from the coast originated by the effect of the northern
winds on the African coast. In addition, we will also eventually
consider vortex trajectories more complex than in \citet{Jung1993}
and \citet{Ziemniak1994}. The kinematic model by Jung et al. was
originally developed to describe the flow behind a cylinder of
radius $r$ located in the middle of a channel of width $W$.
Satisfactory comparison was made with numerical solutions of the
Navier-Stokes equation in the range of Reynolds numbers  such that
the velocity field is periodic in time (von Karman vortex street
flow), i.e., for Re of order $100$. It is remarkable that
numerical simulations in \cite{Aristegui1997} show that already at
Re$\approx 100$, many orders of magnitude smaller than the true
Reynolds number in the turbulent ocean, the flow around Gran
Canaria given by a barotropic quasigeostrophic model reproduces
some of the observed large scale features. This gives confidence
to the hypothesis that the streamfunction in \citet{Jung1993} and
\citet{Ziemniak1994}, developed for flows in that order of
Reynolds numbers, is a good starting point to model the large
scale features of the island wake.

There are, however, many unrealistic features in it. Among them,
the most noticeable is that the true geophysical flow is not time
periodic. Another one is that it lacks of any of the small scale
structures characteristic to real turbulent flows. To minimize
these shortcomings, in this work we will present results for the
transport across the wake of an obstacle for three different
situations: In the first case we will use a streamfunction
periodic in time, which is the direct extension of the model in
\cite{Jung1993} but including an Ekman term. In a second case the
motion of the vortices, which in the original model is
rectilinear, will have a stochastic component, giving rise to a
non-periodic flow. In the third case we will add a random velocity
component to the particle motion in the periodic streamfunction,
as a way to investigate the impact on transport of small-scale
turbulent diffusion. These three situations are described in the
next subsections.

\subsection{Periodic flow}

The spatial coordinates are chosen such that the mean flow runs
along the horizontal $x$ direction, from left to right, put the
center of the cylinder at the origin of coordinates, and measure
lengths in units of the cylinder radius, so that $r=1$. Under
these conditions the streamfunction, based in \cite{Jung1993},
will be written as
\begin{equation}\label{StrfFGH}
\Psi(x,y,t) = f(x,y) g(x,y,t).
\end{equation}
The first factor $f(x,y)$ ensures that the trajectories do not
penetrate into the cylinder,
\begin{equation}\label{StrfF}
f ( x,y )= 1-e^{-a \left(\sqrt{x^{2}+y^{2}}-1 \right)^2} .
\end{equation}

There is a frictional boundary layer of width $a^{-1/2}$ on which
the tangential velocity component tends linearly to zero, while
the radial velocity component decreases quadratically. The
cylinder surface can be considered as the union of an infinite
number of parabolic fixed points.

The second factor $g(x,y,t)$ models the background flow, the
vortices in the wake, and the Ekman term:
\begin{eqnarray}\label{StrfG}
g ( x,y,t )&= & -w h_{1}(t)g_{1}(x,y,t) +w h_{2}(t)g_{2}(x,y,t) \nonumber \\
 &+&u_0 s(x,y) y + u_E (x-1)\Theta (x-1).
\end{eqnarray}
 The first two terms describe the simultaneous presence of
two vortices in the wake. They are of opposite sign but their
maximal vortex strengths are equal and denoted by $w$. They are of
Gaussian shape:
\begin{equation}
\label{Strfgi}
g_{i}(x,y,t)=e^{-\kappa_0 \left[\left(x-x_{i}(t))^{2} +
\alpha(y-y_{i}(t)\right)^{2}\right]} \ , i=1,2
\end{equation}
$\kappa_0^{-1/2}$ is the characteristic linear size of the vortices
(the radius), and $\alpha$ gives the characteristic ratio between
the elongation of the vortices in the $x$ and $y$ direction. The
vortex centers move along the $x$ direction according to
\begin{eqnarray}
x_{1}(t) =& 1+L\left( \frac{t}{T_{c}}\ {\rm mod}\ 1 \right) \ , & y_{1}(t)=y_{0} \nonumber \\
x_{2}(t) =& x_{1}(t-T_{c}/2) \ , & y_{2}(t)=-y_{0} \ ,
\label{Strfxc}
\end{eqnarray}
and their amplitudes are modulated by
\begin{eqnarray}
h_{1}(t) &=&  \left| \sin\left(\pi \frac{t}{T_{c}}\right) \right|  \nonumber \\
h_{2}(t) &=& h_{1}(t-T_{c}/2)
\label{Strfh1}
\end{eqnarray}
Thus, vortices are created behind the cylinder with a dephasing of
half a period. Each of them moves a distance $L$ along the $x$
direction during a time $T_{c}$, then fades out, and the process
restarts.

The third term in $g(x,y,t)$ describes the background flow, a
current of speed $u_0$ in the positive horizontal direction.  The
factor $s(x,y)$ introduces the shielding of this background flow
behind the cylinder, allowing it to be replaced by the vortex
structures. Its precise form is
\begin{equation}
\label{Strfs}
s(x,y)=1-e^{-(x-1)^{2}/\alpha^{2}-y^{2}}.
\end{equation}
This shielded region is of size 1, i.e. of the size of the
cylinder or island.

The last term in $g(x,y,t)$ is absent in the original
streamfunction of Jung et al. \cite{Jung1993,Ziemniak1994}. It
models an additional velocity of constant strength $u_E$ in the
$y$ direction acting only when the $x$ coordinate of a particle is
larger than $1$, i.e. just behind the island ($\Theta$ is the
Heavyside or step function, i.e. $\Theta(u)=1$ if $u>0$ and
$\Theta(u)=0$ if $u<0$). This corresponds to a stream crossing the
vortex street towards the negative $y$ direction just past the
cylinder. This term was introduced in order to take into account
the existence of the Ekman-drift in the region of the Canary
Islands which points towards the ocean interior. Plots of the
streamlines of the flow without and with the  Ekman-drift are
shown in Figs. \ref{fig:strf} and \ref{fig:strf_ek}, respectively.
The rectangle in the upper part is the area where a large number
$N$ of particles, initially equidistant, is repeatedly introduced
at regular time intervals $\Delta$. Their trajectories are
followed by integrating the equations of motion (\ref{Strfcomp})
and from them the cross-wake transport is estimated (see below).
This configuration aims at representing the transport of water
parcels, rich in nutrients, from an upwelling region in the
African coast towards the ocean interior.

\subsection{Non-periodic flow}

Real oceanic flows are never perfectly periodic. It is well known
that structures that are perfect barriers to transport
 (Kolmogorov-Arnold-Moser (KAM) tori) in
a periodic flow become leaky when the time-dependence is not
exactly periodic \cite{Wiggins1992}, so that there is the
possibility that the model defined in the previous subsection
would underestimate transport. In addition, in the above presented
periodic flow case, the trajectories of the vortices are
rectilinear and regular which does not happen in the real case of
Canary vortices. As a way to relax both limitations, we add some
randomness to the vortex trajectories. Instead of moving along
straight horizontal lines, $y_1(t)=y_0$, $y_2(t)=-y_0$, the
vertical coordinates of the vortices move according to
$y_1(t)=y_0+\gamma \xi(t)$, and $y_2(t)=-y_1(t)$, where $\xi(t)$
is a normalized Gaussian white noise ($<\xi(t)>=0$, $<\xi(t)
\xi(t')>=\delta(t-t')$) and $\gamma$ the noise strength. Using
this approach the periodicity of the streamfunction is broken, and
some of the characteristic features of periodic flows, such as the
existence of strict barriers to transport, will not be present in
this case. Again, particle trajectories starting in the upper
rectangle are determined from equations (\ref{Strfcomp}).

\subsection{Periodic flow with turbulent diffusion of the particles}

For the preceding two cases, periodic and non-periodic flows, the
trajectories of tracers are computed by integrating equations
(\ref{Strfcomp}) with the given streamfunction. This
streamfunction contains only large scale features and completely
misses all the small scale turbulence that is characteristic to
the real ocean.

A convenient way to include unresolved small scales in Lagrangian
computations is to add to the velocity field experienced by the
Lagrangian particle a fluctuating term representing small-scale
turbulence \cite{Griffa96,Mariano}. In our case,
\begin{equation}
{\bf \dot x} (t)={\bf v}({\bf x},t)+\sqrt{2 K}{\mathbf{\eta}}(t).
\label{Strfcompturbul}
\end{equation}
where ${\bf x} =(x, y)$ and ${\bf v}({\bf x},t)$ is the velocity
field given by Eq.(\ref{Strfcomp}) for the periodic flow case.
This gives additional diffusion to particle trajectories. Usually
the two-dimensional vector ${\mathbf{\eta}}(t)$ is taken to be a
Gaussian Markov process with a memory time of the order of some
days \cite{Buffoni,Falco2000}. Here, to explore the impact of
irregular unresolved motions in the opposite extreme to the
deterministic situation considered in the previous sections, we
use for ${\bf \eta}(t)$ a two-dimensional Gaussian white noise of
zero mean and correlations $\left<{\mathbf{\eta}}(t) \cdot
{\mathbf{\eta}}(t')\right>=\delta(t-t')$.
 For the strength $K$ we take $K \approx 10\
m^2s^{-1}$, which is the effective eddy diffusivity estimated by
\cite{Okubo} as acting at the spatial scales of about $10\ km$,
which are of the order of the spatial structures that begin to be
missed from our streamfunction. In this case, some typical
features of the periodic flow are lost, and even smooth dynamical
systems structures become fuzzier. In particular, transport may
occur even across perfect Lagrangian barriers.

\section{Parameter estimation for the Canary zone}

In this section we enumerate the relevant geophysical properties
of the upper ocean levels of the Canary zone in order to be
represented in the model. We extract the relevant information from
references \cite{Aristegui1997,Barton1998,Barton2004,Pelegri2005}.
Although there are seasonal variations in most of the parameters,
representative constant values are used here.

\begin{itemize}
\item A unique island is in the model. This mimics the Gran Canaria island,
which seems to have most influence on the zonal mesoscale activity
\cite{Aristegui1997}. Its linear size is of the order of $54\ km$,
from which we take  the radius of the model cylinder to be $r=25\
km$.
This will be taken in the next Sections as the unit of length so
that $r=1$ there.


\item Typically in this area the mean velocity is
$0.05\ m/s$ and in some periods of the year reaches $0.2\ m/s$.
We take a background flow velocity of $u_0=0.18\ m/s$. In
numerical experiments (with a large eddy viscosity)
\cite{Aristegui1997} a von Karman vortex street appears when the
background flow is larger than $0.1\  m/s$.


\item Different sizes are observed for the Canary eddies, ranging from $50$
to $100$ kilometers, and depending on the distance to the island
that generates them. In any case the mean radius of the eddies
$\kappa_{0}^{-1/2}$ is comparable to that of the island that generates
it. Thus we take $\kappa_{0}^{-1/2}=r$.

\item Eddies are usually elliptic. Its eccentricity diminishes
with the distance to the island.
In our model we take $\alpha=1$, that would represent circular
vortices. But due to the part of the streamfunction representing
shielding behind the cylinder, vortices are stretched and have
some ellipticity.


\item The rotation period of a buoy in an eddy \cite{Pelegri2005}
ranges from $3$ to $6$ days (although increasing with time). This
gives a linear velocity at their periphery (distance $r$ from the
center) of about $0.6\ m/s$. By equating this speed with typical
values of derivatives of the streamfunction at the vortex
periphery we estimate the vortex strength $w \approx 55 \times
10^3\ m^2/s$.

\item The shedding of eddies, as already  commented,
 is not perfectly periodic. Nevertheless
we take a typical interval between eddy shedding events of $15$
days. Thus $T_c=30$ days.

\item Some measurements of eddy velocities indicate that they move towards
the southwest at a velocity of $5-6$ kilometers per day. The
typical displacement during a time $T_c$ is thus $L=150\ km=6 r$.

\item Lifetime of the eddies is of several weeks,
with some measurements reporting lifetimes of several months.
Typically they remain close to the island for about one week. In
the model the lifetime is the same as the period $T_c$, which is
within the order of the magnitude of observed permanence in the
wake.

\item $a^{-1/2}$, the width of the
 cylinder boundary layer, is difficult to estimate
since it is ill-defined at geophysical scales. Fortunately its
value only affects motion close to the cylinder and its effect is
unimportant in most of the velocity field.
We take $a^{-1/2}=r$.

\item The Ekman flow is originated by the wind stress
$\tau= \rho_{air}c_d v^2$, where $\rho_{air}=1.222\ kg/m^3$ is the
air density, $c_d \approx 0.0013$ is the drag coefficient between
water and air, and $v$ is the wind velocity, typically in the
range $[2.7, 8.7]\ m/s$. Thus the wind stress is in a range $[0.012,
0.12]\ N/m^2$. For a particular intermediate value of $v=5\ m/s$ we
have $\tau=0.040\ N/m^2$.

The value of the Ekman speed is given by:
\begin{equation}
u_E= \frac{\tau}{\rho_0 f h},
\end{equation}
being $\rho_0 \approx 1024\ kg/m^3$ the sea water density,
$f=10^{-4}\ s^{-1}$ the Coriolis parameter at the Canary latitude,
and $h$ the depth
of the Ekman layer. It ranges between $15\ m$ and $100\ m$. We
take the intermediate value $h=50\ m$, which can be justified from
the expression $h= \sqrt{\frac{2 A_v}{f}}$ for a vertical
turbulent viscosity $A_v \approx 0.1\ m^2/s$. With these parameter
values, and the values for the wind stress, the Ekman velocity
$u_E$ is in the range $[0.0023, 0.02]\ m/s$.

\end{itemize}

In the following, in addition to measure lengths in units of $r$,
we measure time in units of $T_c$. With this, the non-dimensional
values of the parameters to be used in the model read:
$r=T_c=a=\alpha=1$, $\kappa_0=1$, $L=6$, $u_0=18.66$,
 $u_E \in [0.2, 2]$,
$w=200$. In the non-periodic case we use $\gamma=0.5$ and
$y_0=0.5$ (i.e. half the island radius) for the parameters of the
vortex trajectories.

\section{Particle dynamics in the wake}

The dynamics given by Equations~(\ref{Strfcomp}) can be
interpreted as the equation of motion of a one-degree of freedom
Hamiltonian system with time-dependent Hamiltonian $\Psi(x,y,t)$.
In our case we have an open system, meaning that the particles
start in an incoming asymptotic region, pass a region where the
dynamics is time dependent, and then leave the system through an
outgoing asymptotic region. While in the time-dependent region of
the system, particles are trapped by the vortices and whirled
around for a while. Since the velocity field is time dependent,
particles can be handed from one vortex to the following one and
can remain in the region close to the cylinder for a relatively
long time, even though the vortices leave this region quite
rapidly. In fact there are periodic and localized trajectories
organizing these long scattering orbits. They constitute the
backbone of the so-called {\sl chaotic saddle}, the unstable set
of trajectories never leaving the wake region \citep{Jung1993}.
This structure, and particularly its stable and unstable manifolds
(the lines along which particles ending at the saddle approach it,
and particles close to the saddle leave it, respectively) organize
important trajectory characteristics in the wake. 
 More in detail, the stable or
contracting manifold of the chaotic saddle is the set of spatial
points $x$ such that particles starting from $x$ approach the
chaotic saddle as time advances. Similarly, the unstable or
stretching manifold is the set of points such that their
backward-in-time evolution approaches the chaotic saddle.
Stable manifolds cannot intersect with themselves
and with other stable manifolds, and the same
holds for the unstable manifolds.
 Moreover, particle trajectories cannot cross these manifolds. However,
stable and unstable manifolds can intersect each other.
 All these
properties make them important templates organizing the particle
trajectories in the system. Typically, vortex boundaries are areas
of tangencies between stable and unstable manifolds. In the
particular case of open flows, like the one studied in this work,
the unstable manifold of the chaotic saddle is the set of points
along which particles leave it, and, therefore, it is the set that
is traced by a number of particles when they are launched in the
system and take an long time to abandon it.

 In Fig.
\ref{fig:manifolds} we show the stable and unstable manifolds of
the chaotic saddle for two representative parameter sets. A clear
change in the shape of the unstable manifold is seen when
increasing the vortex strength $w$. The chaotic saddle itself,
however, is only approached by particles starting on its stable
manifold. In our configuration, and for the parameter values we
use, these structures are closely packed very near the cylinder
surface (see Fig. \ref{fig:manifolds}), in contrast with other
situations studied in the literature
\citep{Jung1993,Ziemniak1994}. As a consequence, for the initial
conditions to be used in Sect. \ref{sect:quantifying}, tracked
fluid particles will not intersect such manifold and they will not
follow strongly chaotic recirculating orbits, but rather they will
be advected downstream relatively fast. In addition they will
leave the wake region along paths that are not perfectly aligned
with the saddle's unstable manifold. Despite of this, we will see
that the change of topology observed in Fig. \ref{fig:manifolds}
has global consequences in particle dynamics and that for
realistic parameter values, corresponding to the right panel of
Fig. \ref{fig:manifolds}, particles can cross the wake, and
experience some stretching and dispersion.

\section{Quantifying transport across the wake of an island}
\label{sect:quantifying}

In this Section we report the numerical results obtained for the
three different flows introduced in Section 2. Our objective is to
quantify transport across a vortex street in the presence of a
continuous source of particles representing the water parcels
upwelling at the African coast. We model this source by placing
test particles in the rectangle $0<x<1$ and $2.1<y<2.5$, i.e.,
above and at some distance of the cylinder (see Figs.
\ref{fig:strf} and \ref{fig:strf_ek}). We place 200 new particles
in the rectangle at regular intervals of time $\Delta=0.01$ (in
units of the flow period $T_c$), i.e. 20,000 new particles per
period, and integrate their evolution under the flow. Particles
are initially placed along four horizontal lines inside the
rectangle, but this has no influence on the results described
below.

Our open system is considered to be the region displayed in Figs.
\ref{fig:strf} and \ref{fig:strf_ek}. Trajectories leaving this
region are no longer integrated. The idea is that it will be
impossible (for the periodic flow case) or practically impossible
(for the non-periodic and turbulent cases) for the particles to
return back to the region after they leave it. We count at every
interval of time $\Delta$ how many particles have {\sl crossed the
wake}. There is some ambiguity in defining the transverse extent
of the wake. Fortunately, as will become clear from the results
presented later, the dynamics is such that particles either do not
approach the central region behind the cylinder while they remain
inside our region or rather they perform a rather large transverse
excursion. Thus, any reasonable definition of `crossing the wake'
will give essentially the same results. In this Section we will
count particles crossing the central line $y=0$ as `having crossed
the wake', and at the end of the Section we will show that the
same results are obtained if counting them when crossing $y=-1$
(see Fig. \ref{fig:comparacionY}). Particles crossing the chosen
line several times are counted only once.

Because of the presence of the Ekman term $u_E$, all trajectories
will eventually reach arbitrarily negative $y$ coordinates if
observed sufficiently far downstream. Clearly, this can not be
considered to be a wake crossing, and we restrict our computation
to the region shown in Figs. \ref{fig:strf} and \ref{fig:strf_ek},
where the vortices remain localized and thus it is the only part
of the flow in which nontrivial dynamics occurs. Given the simple
structure of the flow, even when wake crossing occurs, the
particles can not go very far and typically they will not leave
the proximity of the wake region. In a more realistic ocean setting,
additional mechanisms can occur after wake crossing that may bring
particles further towards the open ocean. But wake crossing is
anyway the first step needed for such long-range transport to occur.

We fix all but two of the parameters of the model indicated in the
previous section, namely the strength of the vortices, $w$, and
the Ekman pumping $u_E$, which are varied in a realistic range.
The measure of transport across the wake is performed by counting
the number of particles crossing the line $y=0$ during each time
interval $\Delta$. A short transient after the launching of the
first particles in the rectangle this quantity becomes a periodic
function of time in the periodic flow case, and approximately
periodic under the other two flows. To focus on average transverse
transport, a quantity called $N^c$ is computed as the ratio
between the number of particles crossing $y=0$ during 6 flow
periods (after discarding an initial transient of 3 periods) and
the total number of particles launched during that time (120,000
particles).


In Fig.~\ref{fig:perw} we plot $N^c$ versus $w$ for different
values of the Ekman pumping strength $u_E$ in the periodic flow
case. The most relevant result is the absence of transport for
small values of $w$, identifying the existence of a barrier that
does not permit the entrance of particles in the wake. This
barrier disappears when $w$ crosses a critical value $w_c$ which
depends on the Ekman pumping strength $u_E$, $w_c=w_c(u_E)$. At a
fixed value of $u_E$ the proportion of crossing particles
increases with increasing $w$ as expected, above the critical
threshold $w_c$. In Fig.~\ref{fig:perwcrit} we plot
the critical value $w_c$ as a function of $u_E$. The value of
$w_c$ diminishes from approximately $w_c=50$ for $u_E=0.2$ to
$w_c=20$ for $u_E=2$. Similarly for increasing $u_{E}$ the
ratio of particles crossing is significantly larger for a
fixed $w$. Most importantly, for the typical realistic value
$w=200$ (see Sect. 3) one obtains a rather large proportion of
crossing particles, $N^c$, independently of the value of $u_E$,
so that one can
expect that behind the Canary islands a net transport of particles
from the coast to the opposite side of the island wake occurs.

Now we illustrate the transport mechanism by looking at particle
distributions for the two different situations just identified. In
Fig.~\ref{fig:tracersperi} we show a snapshot of the particle
positions, all of them launched at the horizontal lines in the
marked rectangle at successive times. In the left panel we plot
the case $w=10$ in which the launching site is on the exterior of
the barrier impeding transport across the wake. In the right panel
($w=200$) we observe particles spreading through the wake of the
island. Obviously a barrier no longer exists between the launching
site and the lower parts of the wake. Transport occurs along a
filament entrained into the wake that stretches the particle lines
and later disperses them. The similarity of the tracers
distribution with real features observed in the Canary area is
remarkable (see for example Fig. 24 of \cite{Barton1998}, or Fig.
1 of \cite{Aristegui2004}). This occurs despite the fact that, as
advanced before, the trajectory structure in Fig.
~\ref{fig:tracersperi} is rather different from the saddle
manifolds in Fig.~\ref{fig:manifolds}. The difference arise
because the launching site for the tracer particles is rather far
from the manifolds.

Thus we can conclude that the formation of vortices in the wake of
the Canary Islands together with Ekman pumping make up a possible
mechanism for the formation and entrainment of nutrient-rich
filaments into the Canary wake, and eventually for transport of
nutrients from the African coast to areas in the Atlantic beyond
the islands. The effectiveness of this enrichment mechanism will
depend however on how fast are the upwelled nutrients consumed by
the biological populations near the coast \citep{Pelegri2005}.
 Note that since the size of the island (and
so the width of the wake) is $50 \ km$, the time
the particles would need to cross the wake if driven only by the
Ekman flow ($u_E$ is in the range $[0.0023, 0.02]\ m/s$) is between
$28-250$ days. Given that the mean flow $u_0=0.18 \ m/s$
transports particles out of the observation region shown in the
figures ($10$ cylinder radii) in about $16$ days, we see that no
particles are able to cross the wake region behind the cylinder
under the sole effect of the Ekman flow. 

We now discuss the results for the other two flows considered. In
Fig.~\ref{fig:aperw} we show $N^c$ versus $w$ for different
$u_{E}$ in the non-periodic flow case, that is, with the random
$y$ component for the trajectories of vortex centers. The
non-periodicity of the flow has been introduced to overcome strict
barriers to transport that are not realistic. The results show
that the effect of non-periodicity is not strong. We still observe
a value $w_c$ below which transport is extremely low, though it is
non-zero now. Some particles can enter into the wake at low vortex
strength $w < w_c$ due to the non-periodic nature of the flow, but
their number is rather small. For higher vortex strength $w>w_c$
we observe an increasing net transport with increasing $w$. For a
fixed $u_E$ the critical value $w_c$ is lower than the
corresponding threshold in the periodic case.
Fig.~\ref{fig:tracersnperi} shows distributions of tracers, again
for $w=10$ and $w=200$. They share the qualitative features with
the periodic case, Fig.~\ref{fig:tracersperi}, although now there
is much more particle dispersion after the filament enters the
wake.

The same plots for the case of particles driven by the periodic
flow with turbulent diffusion are shown in Figs.~\ref{fig:turw}
and \ref{fig:tracersperiturb}. For small values of the Ekman
pumping $u_E$ there is again a critical vortex strength $w_c$ such
that for $w < w_c$ only minimal transverse transport is observed.
In the realistic value, $u_E=2$, a non-negligible net transport of
particles is observed already for low vortex strength $w$, in
contrast to the previous cases where below the critical value
$w_c$ transport was very low. Nevertheless there is still a sharp
increase in effective transport when increasing w. Thus a remnant
of the critical value $w_c$ is still visible. In
Fig.~\ref{fig:tracersperiturb} the distribution of tracers is
plotted for $w=10$ and $w=200$, as in the previous cases. As expected
from the introduction of turbulent diffusion particles become
randomly dispersed, but always around average paths similar to the
previous cases.

The comparison among the three cases (periodic, non-periodic and
turbulent) is shown in Fig.~\ref{fig:comparacion}. Here we fix
$u_{E}=0$, $u_{E}=1$ and $u_{E}=2$, and plot $N^c$ vs $w$ for the
different types of flows in one graph.

The smallest transverse transport, and the largest critical $w_c$,
is always attained in the periodic flow case (which is the only
case for which below $w_c$ transverse transport is exactly zero).
The smallest effective value of $w_c$ is found for the case of
particles with turbulent diffusion, and transport is also higher
in this case for the smallest values of $w$ and all $u_E$. At the
other end of the considered range of $w$, i.e. going towards
realistic values $w>150$, the measured transport $N^c$ is largest
for the non-periodic flow case. The addition of the turbulent
particle diffusion slightly increases transport at large $w$ with
respect to the purely periodic case, but the difference between
these two cases is not large. This indicates that turbulent
diffusion, at least as modelled here, has no strong influence on
transverse transport in the realistic limit of large $w$, while
non-periodic vortex movement is more significant.

Finally we return to the question about the distance over
which transport across the wake occurs by comparing $N^c$ for
different positions of the line which the tracers have to pass in
order to be counted as particles that have actually crossed the
wake. In Fig.~\ref{fig:comparacionY} we compare the proportions of
particles crossing the lines $y=0$ and $y=-1$ for the periodic
flow case and $u_E=1$. Similar results are obtained for other
values of $u_E$ and for the other two flows. The ratio of
particles crossing the line $y=0$ is slightly higher, but both
remain very similar over the entire range of $w$, meaning that the
measured transport does not depend significatively on the choice
of the position of the line the tracers must cross in the wake.
This was in fact quite obvious from the shape of the particle
distributions (Figs. \ref{fig:tracersperi},\ref{fig:tracersnperi}
and \ref{fig:tracersperiturb}).

\section{Conclusions}

The biological activity around the Canary Islands and in the open
ocean depends crucially on the availability of nutrients. An
important source of these nutrients is provided by the upwelling
near the African coast. In a simple scenario, we have shown that
these nutrients can be transported over long distances
perpendicular to the coast due to the formation of filaments that
are entrained into the Canary wake by the eddies present there.
The intensity of this horizontal transport depends strongly on the
vorticity content of the vortices, characterized by $w$, and the
strength of the Ekman pumping $u_E$. Other parameters do not
affect transverse transport so strongly, except the mean flow
velocity $u_0$ that has the obvious effect of transporting
the particles faster or slower downstream the wake, and thus decreasing
or increasing, respectively, their chances to cross it. The
mesoscale structure mediating the crossing -- a meandering
filament -- is very similar to real structures observed in the
Canary region.

 Since our approach is kinematic, and parameters are directly
obtained from observations, our conclusions do not depend on the
particular mechanism producing the wake eddies, being it flow
separation, wind stress curl on the lee of the islands
\citep{Aristegui1997,Barton1998}, etc. The simplicity of our
approach has allowed us to identify different factors which
enhance or diminish transport across the wake. We have found that
cross-wake transport occurs always at large enough $w$. Periodic
flow contains transport barriers which block the transport from
outside the wake if the vortex strength is below a critical
threshold value. Such a sharp threshold is replaced by a crossover
to low transport in the non-periodic case, or when a simple model
of turbulent diffusion is considered. These last mechanisms
enhance transport, but in rather different ways: While turbulent
diffusion influences transport for small vortex strength, mainly
below and close to the critical transport threshold,
non-periodicity of the flow enhances transverse transport at high
vortex strength. Our model of turbulent diffusion acts effectively
only at small scales, while non-periodicity changes the flow on
larger scales yielding a stronger overall effect.

Here we have considered only transport of passive tracers. To
study the impact of the phenomena discussed here on the biological
productivity off from the African coast, coupling to models of
plankton dynamics should be performed. This remains an interesting
task for the future. {\bf Probably, vertical upwelling produced
inside the cyclonic eddies would have to be taken into account}.

{\sl Acknowledgements.} We would like to thank Tam\'as T{\'e}l for
many valuable discussions. M.S. and U.F. would like to acknowledge
the financial
 support by the DFG grant FE 359/7-1(2003).
E.H-G. and C.L. acknowledge financial support from MEC (Spain) and
FEDER through project CONOCE2 (FIS2004-00953). Both groups have
benefited from a MEC-DAAD joint program. C.L. is a {\sl Ram{\'o}n
y Cajal} fellow of the Spanish MEC.

\clearpage

FIGURE LEGENDS

Fig. 1. The Canary Islands region, with the Canary current
running southwestwards parallel to the African coast, where there
is an intense upwelling, and impinging on the islands.

Fig. 2. The streamlines of the flow without Ekman flow,
$u_{E}=0$, at vortex strength $w=200$. Other parameters as
described in Sect.~3. The box where tracers are starting is drawn
just above the cylinder with coordinates $0 < x < 1$, and $2.1 < y
< 2.5$. The snapshots are at $t=0$ (top-left), $t=T_{c}/4$
(top-right), $t=2T_{c}/4$ (bottom-left), and $t=3T_{c}/4$
(bottom-right).

\begin{figure} \centering{
\mbox{
\includegraphics [angle=270, width=0.5\textwidth]{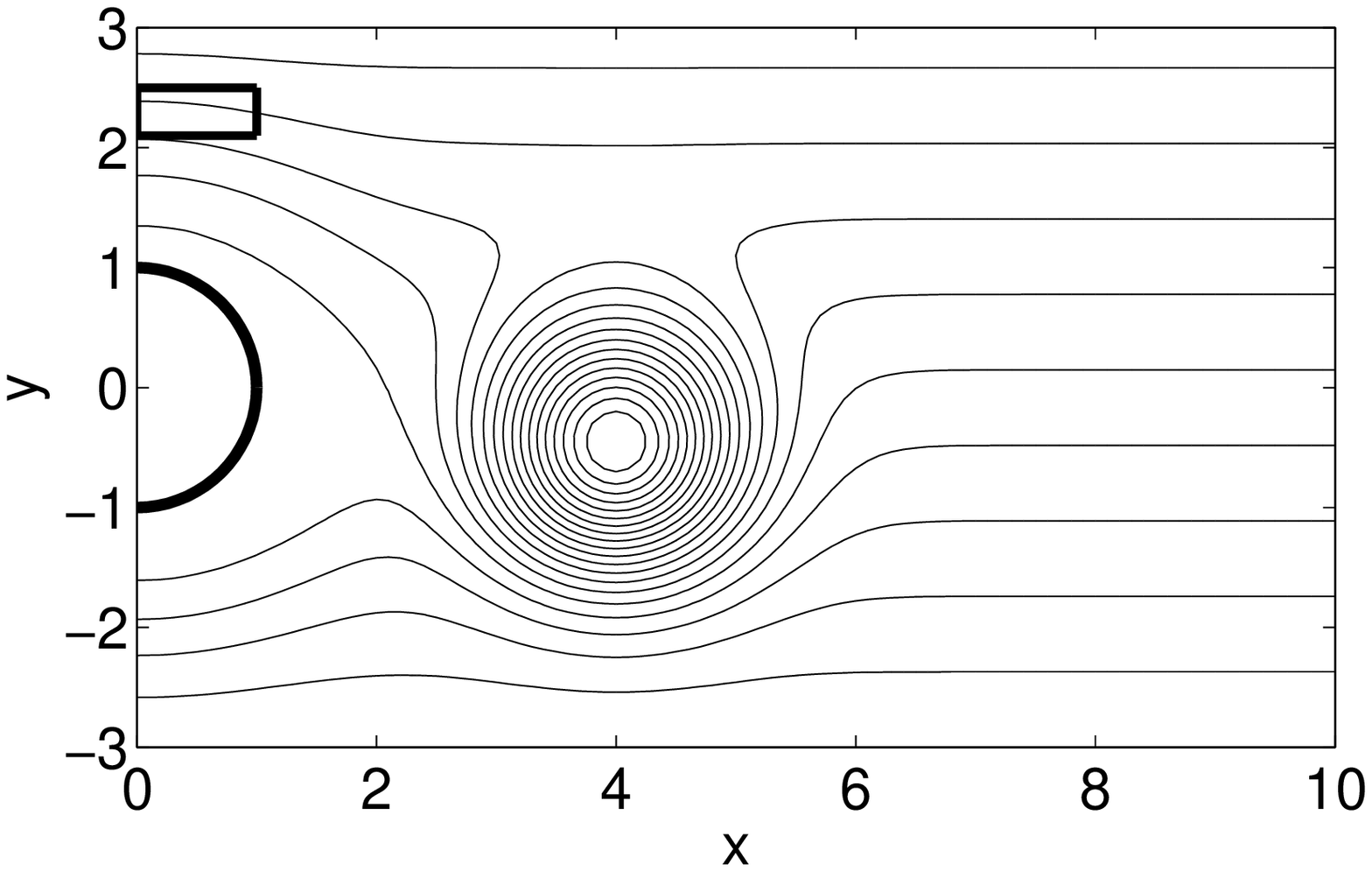}
\includegraphics [angle=270, width=0.5\textwidth]{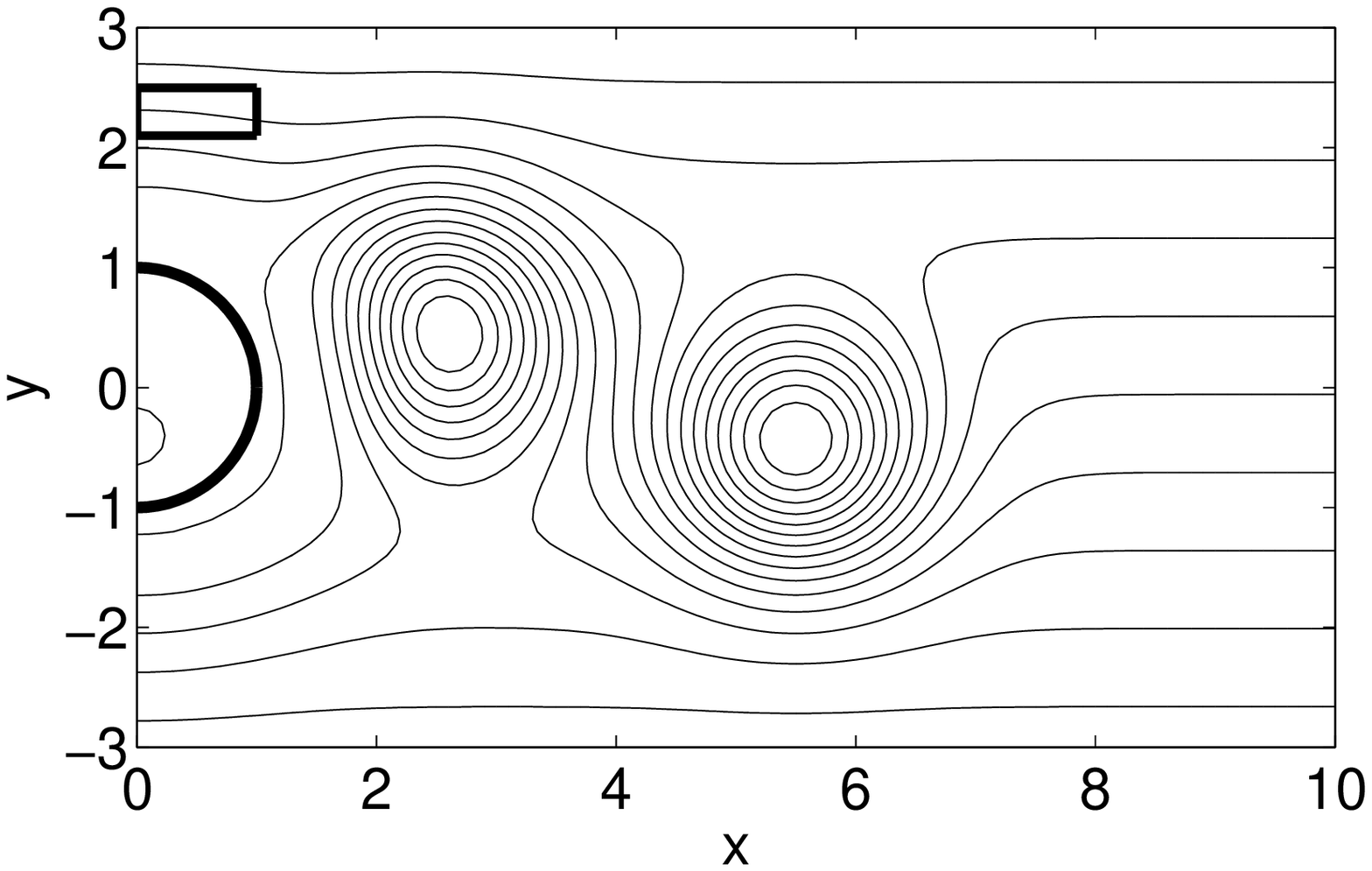}
}
\mbox{
\includegraphics [angle=270, width=0.5\textwidth]{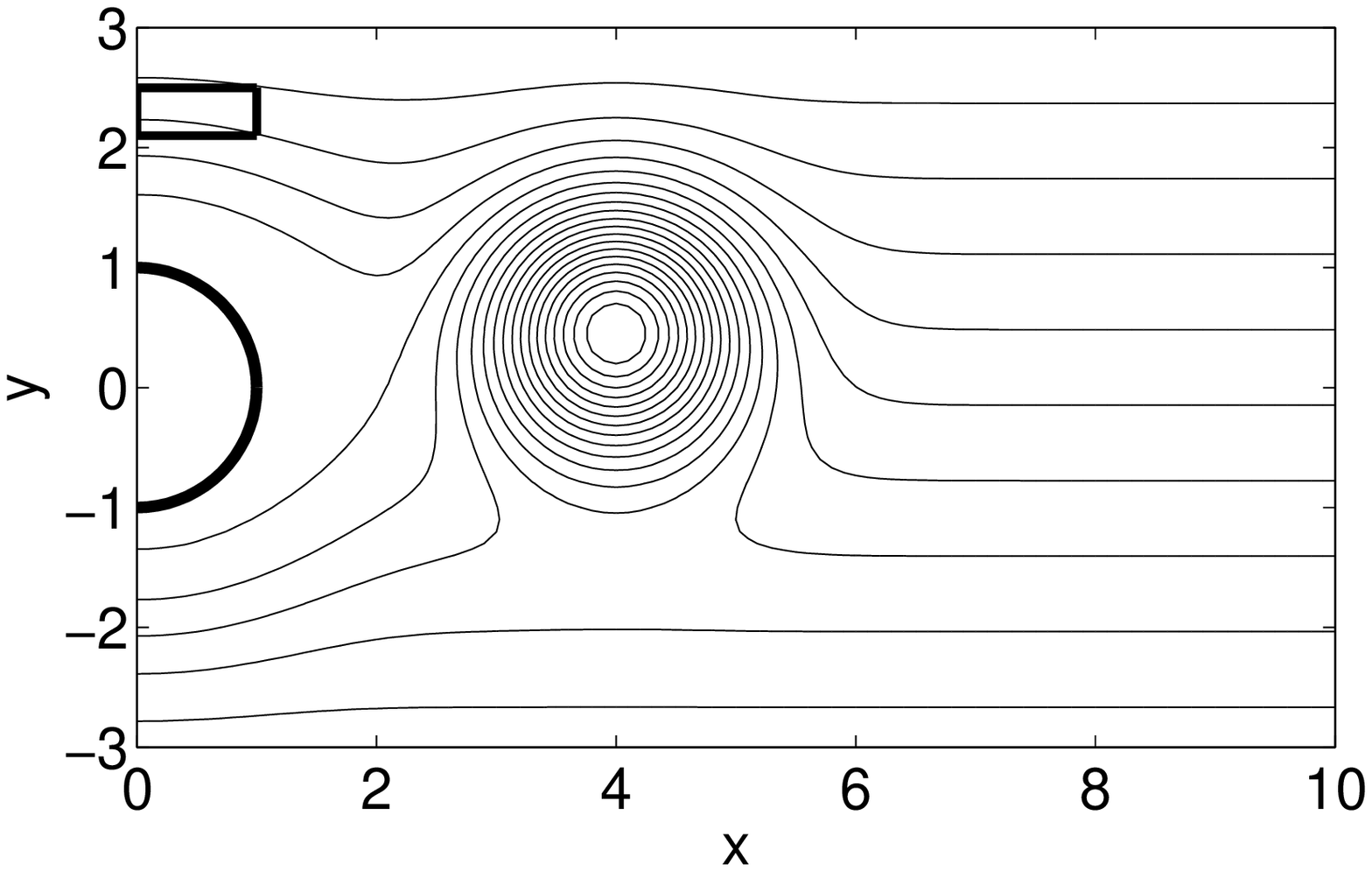}
\includegraphics [angle=270, width=0.5\textwidth]{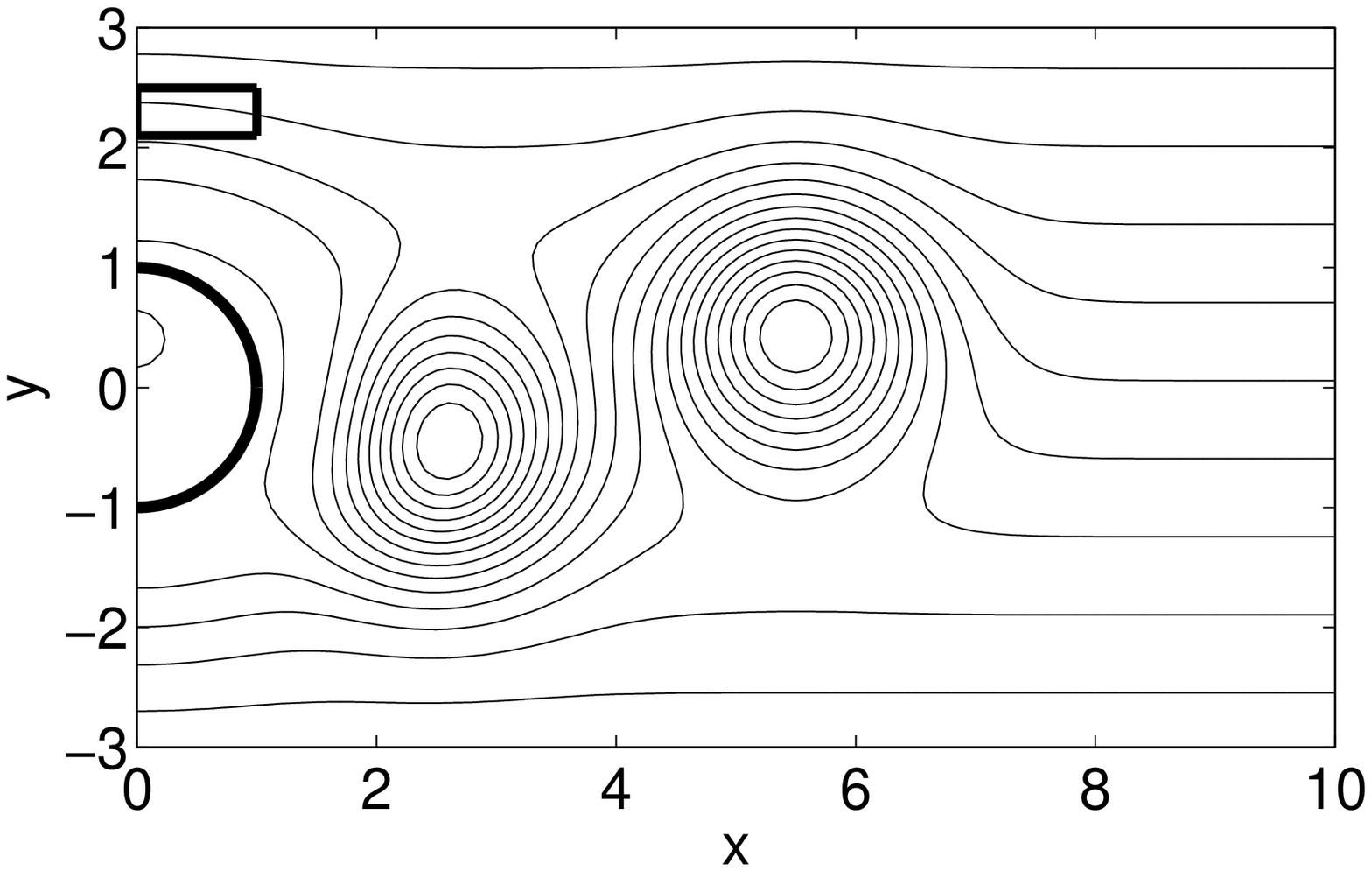}
} }\caption{The streamlines of the flow without Ekman flow,
$u_{E}=0$, at vortex strength $w=200$. Other parameters as
described in Sect.~3. The box where tracers are starting is drawn
just above the cylinder with coordinates $0 < x < 1$, and $2.1 < y
< 2.5$. The snapshots are at $t=0$ (top-left), $t=T_{c}/4$
(top-right), $t=2T_{c}/4$ (bottom-left), and $t=3T_{c}/4$
(bottom-right).
\label{fig:strf}}
\end{figure}

\begin{figure}  \centering{
\mbox{
\includegraphics [angle=270, width=0.5\textwidth]{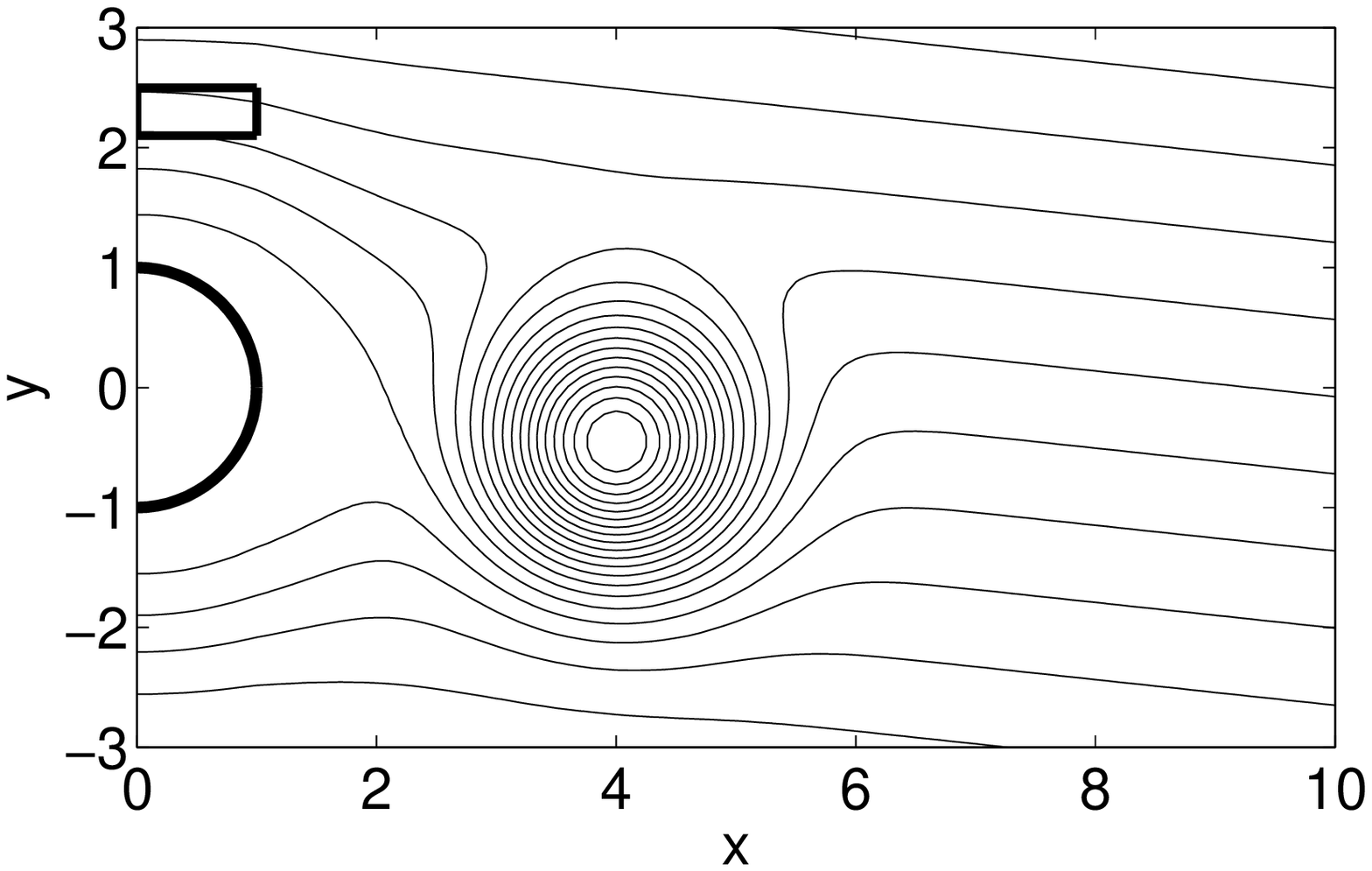}
\includegraphics [angle=270, width=0.5\textwidth]{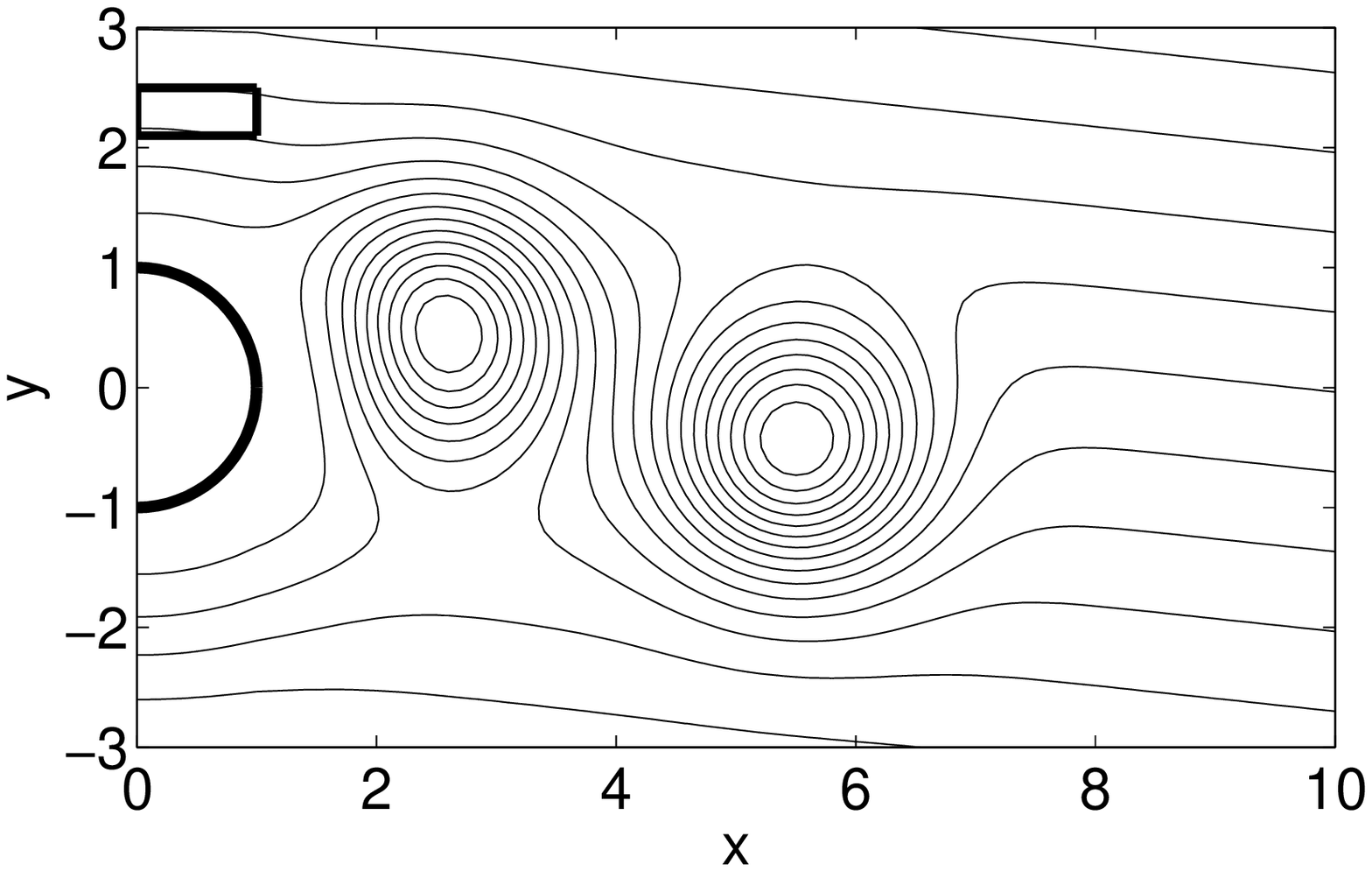}
}
\mbox{
\includegraphics [angle=270, width=0.5\textwidth]{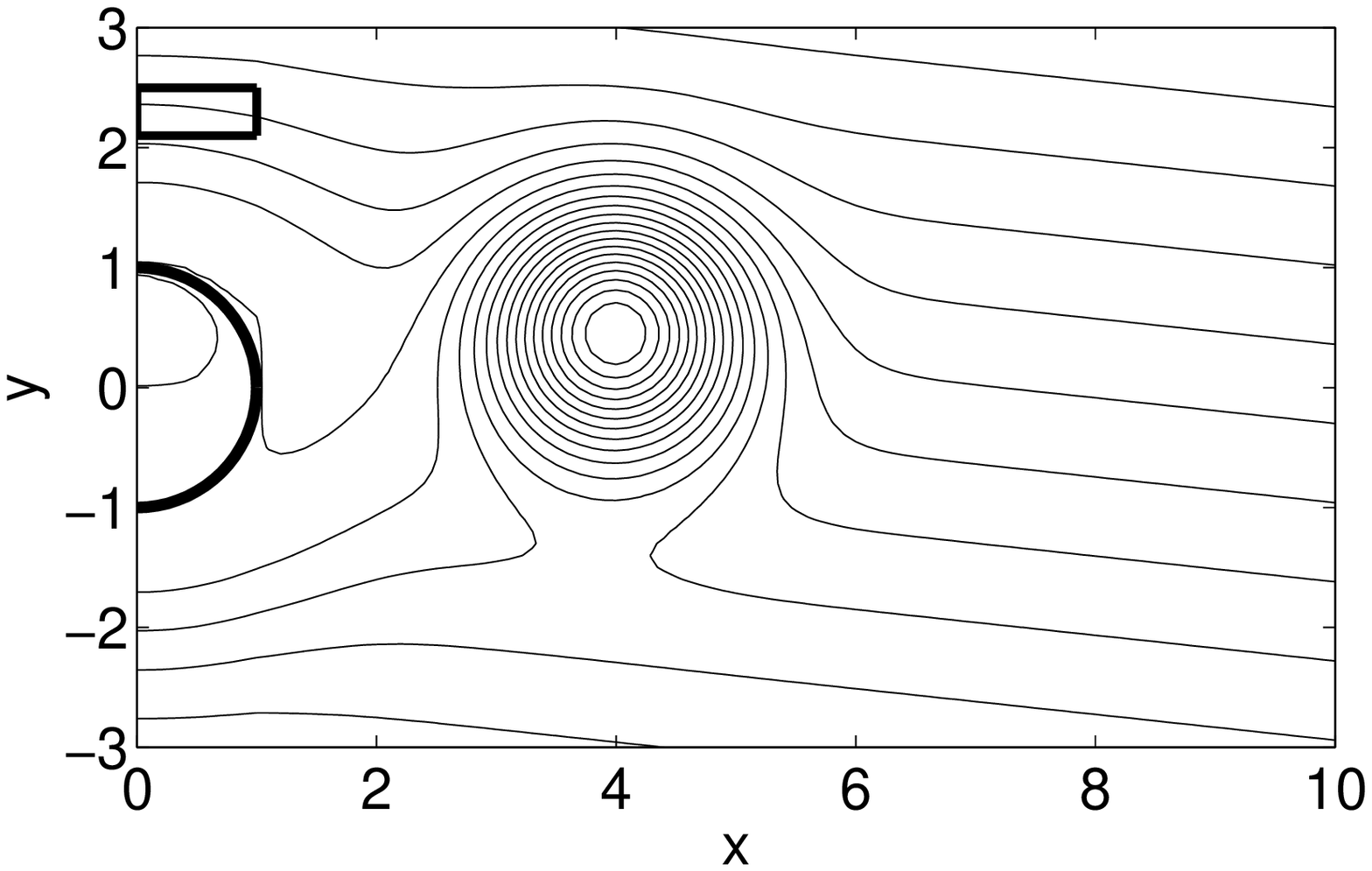}
\includegraphics [angle=270, width=0.5\textwidth]{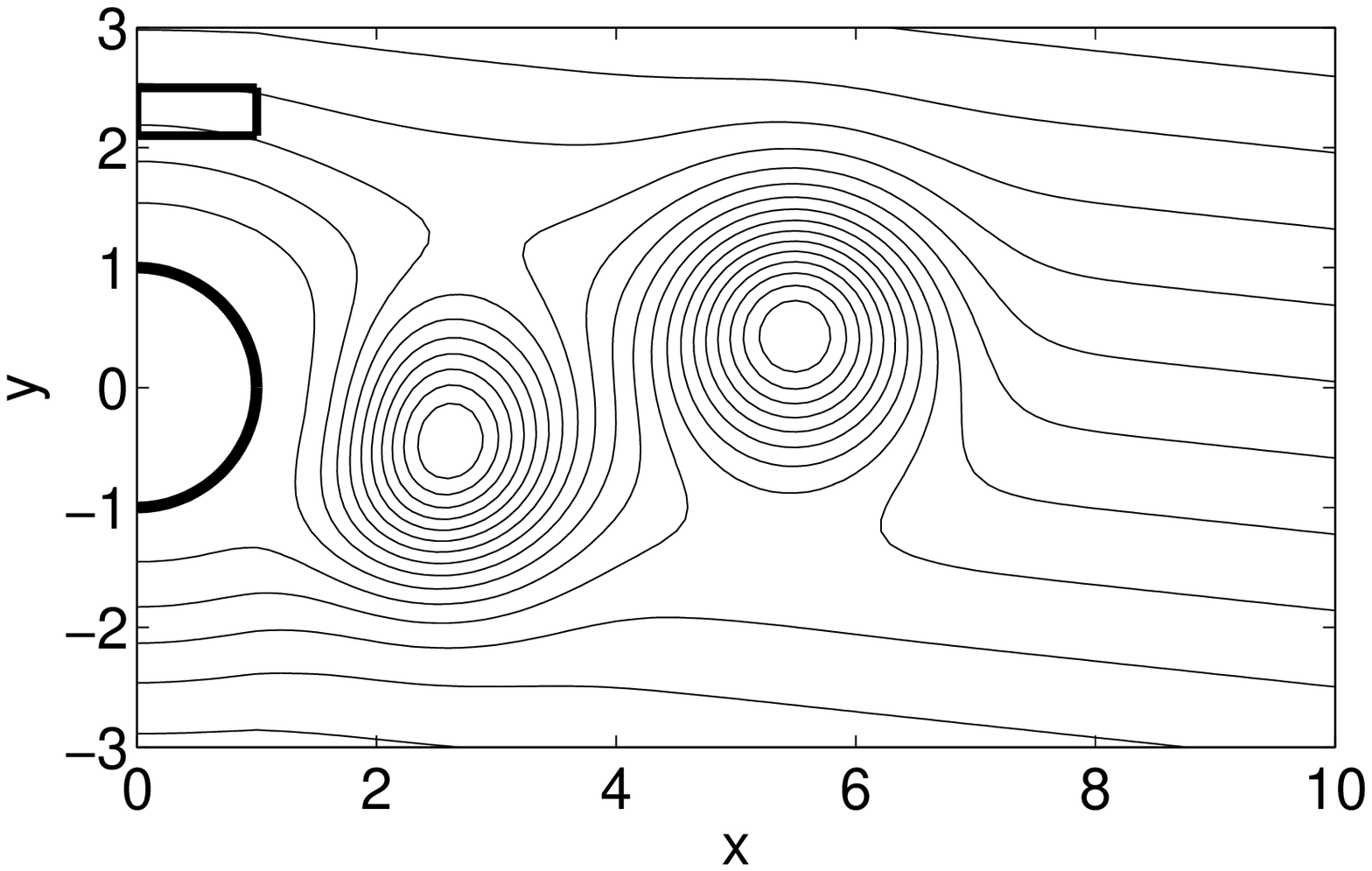}
} }\caption{Streamlines of the flow  with Ekman flow $u_E =2$ and
other parameters, and time sequence of the snapshots, as in Figure
\ref{fig:strf}.
\label{fig:strf_ek}}
\end{figure}

\begin{figure}
\centering{ \mbox{
\includegraphics [width=\textwidth, angle=0]{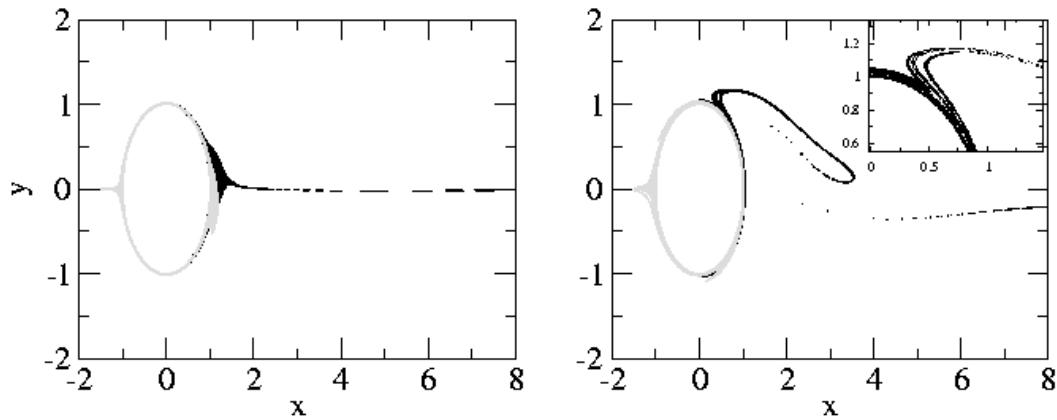}
} }\caption{Stable (in gray) and unstable (in black) manifolds of
the chaotic saddle in the wake of the cylinder, for the case of
the periodic flow and $u_E=2$. Left: Snapshot taken at time
$7T_c$ for vortex strength $w=10$. Right: Snapshot taken at
time $7T_c$ for vortex strength $w=200$. In the inset
we show a zoom of the manifolds in the area close to the cylinder.
 Other parameters as
described in Sect. 3. The chaotic saddle itself is closely packed
immediately behind the cylinder surface. The unstable manifold has
been plotted by releasing a large number of particles left of the
cylinder and very close to it, letting the flow to transport them
for a long time ($7T_c$ as already indicated)
 so that only the ones lasting at this time in the
wake region are still there and plotted. The stable manifold is
plotted in the same way but releasing the particles right of the
cylinder and running the flow backwards in time.
\label{fig:manifolds}}
\end{figure}

\begin{figure}
\includegraphics [angle=0, width=0.8\textwidth]{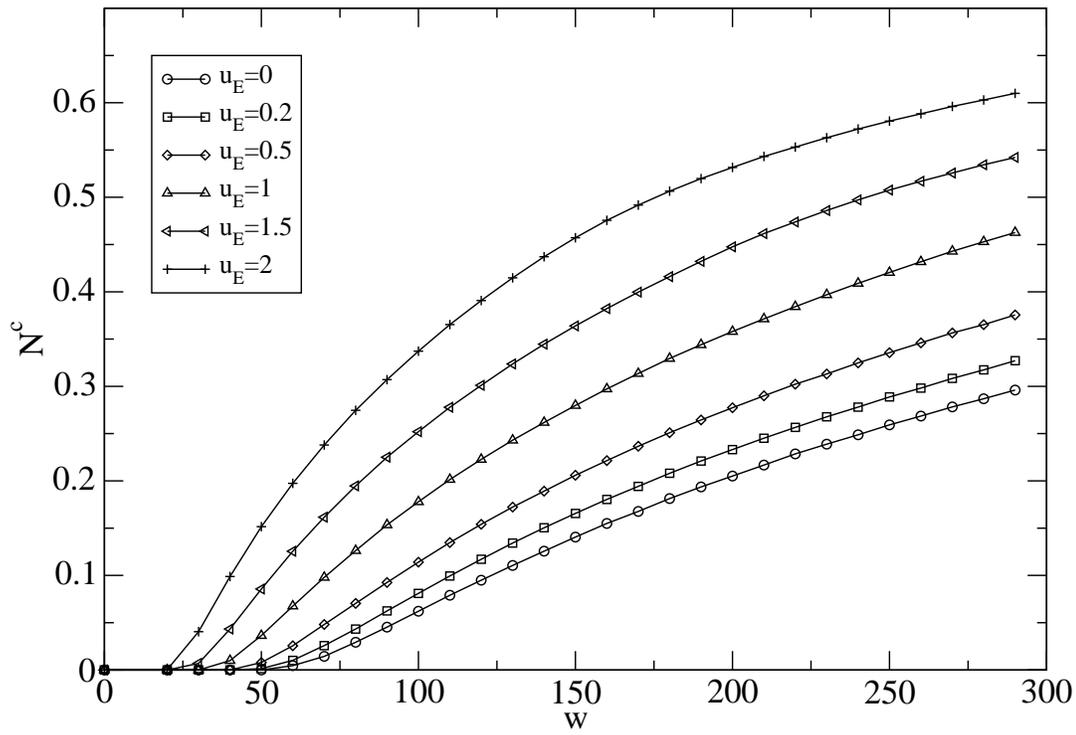}
\caption{The ratio of particles crossing the
wake, $N^c$, versus vortex strength $w$ in the periodic flow. The
different curves correspond to different values of the Ekman
pumping, $u_E$, as indicated in the legend.}
\label{fig:perw}
\end{figure}

\begin{figure}
\includegraphics [angle=270, width=0.8\textwidth]{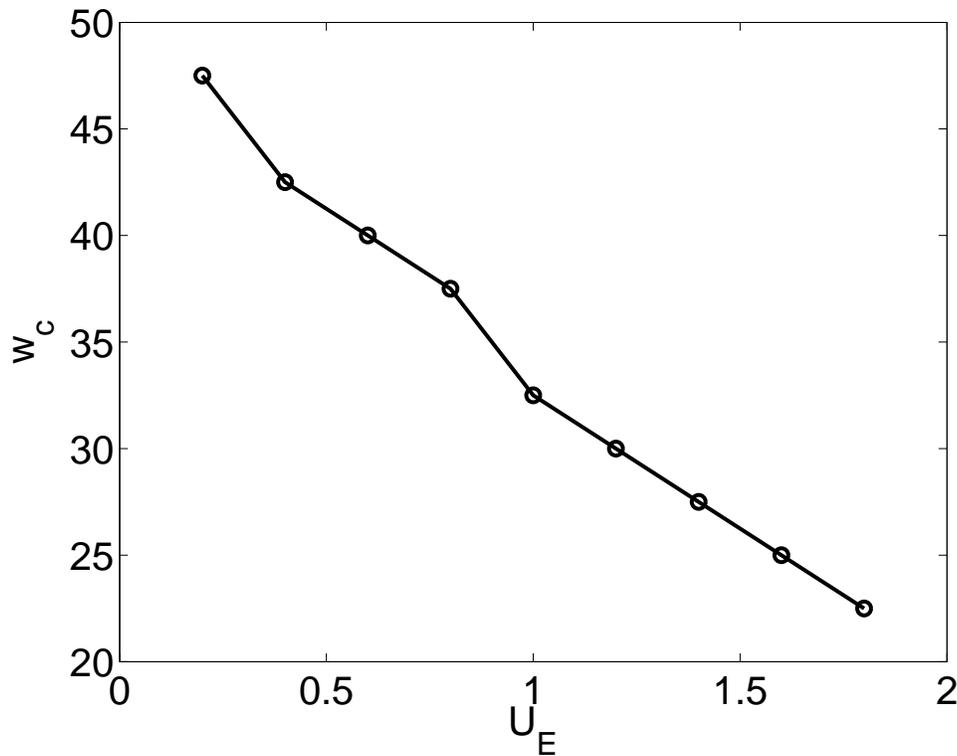}
\caption{The critical values of the vortex strength, $w_c$, versus the velocity
of the Ekman pumping $u_{E}$ in the periodic flow. }
\label{fig:perwcrit}
\end{figure}

\begin{figure}
\centering{
\mbox{
\includegraphics [angle=270, width=0.5\textwidth]{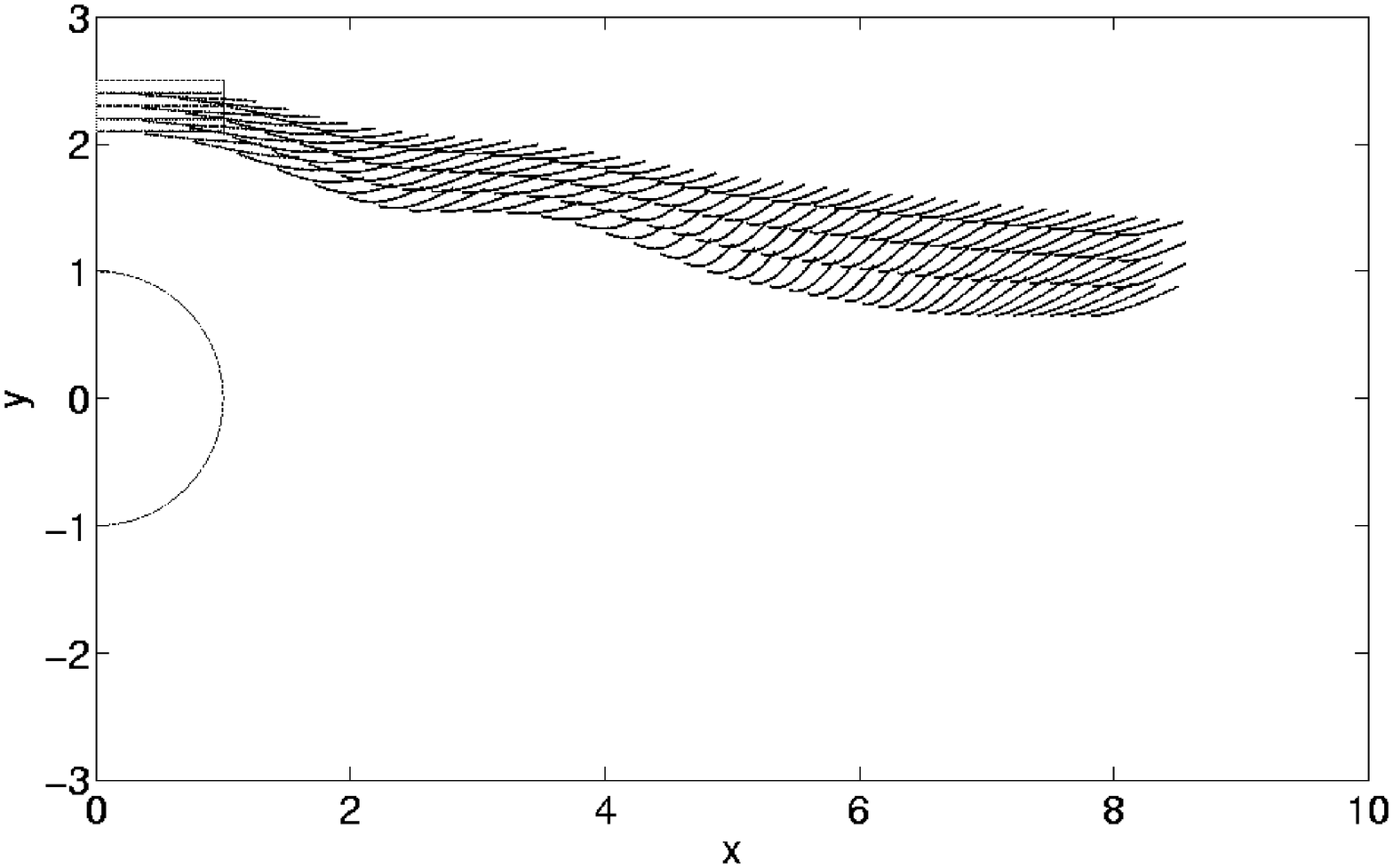}
\includegraphics [angle=270, width=0.5\textwidth]{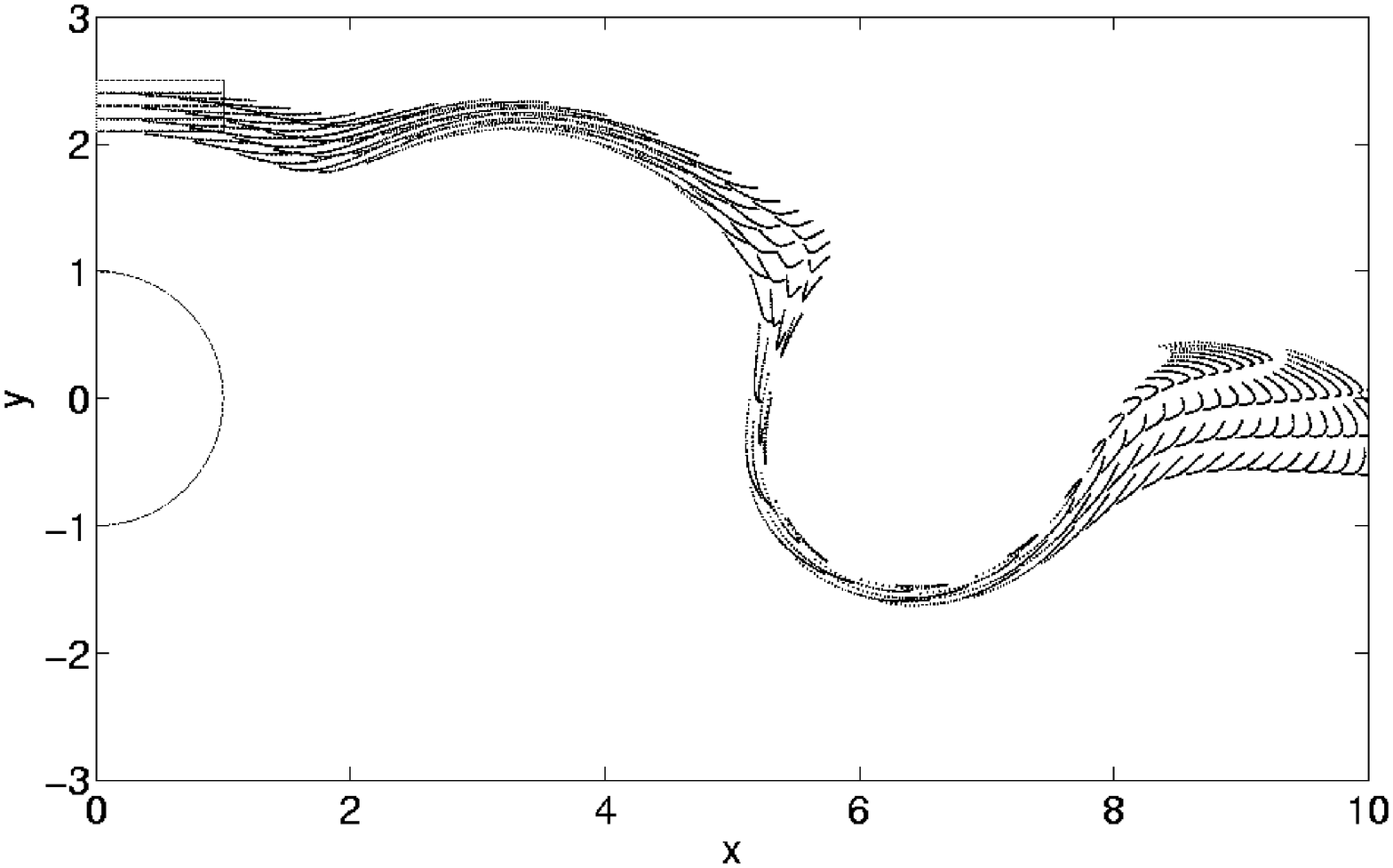}
} }\caption{Plot of the spatial distribution of the tracers in the
wake of the island for the case of the periodic flow and $u_E=2$.
Left: Snapshot of the distribution of the tracers at time
$0.39T_c$ for vortex strength $w=10$. Right: Snapshot of the
distribution of the tracers at time $0.39T_c$ for vortex strength
$w=200$.
\label{fig:tracersperi}}
\end{figure}

\begin{figure}
\includegraphics [angle=0, width=\textwidth]{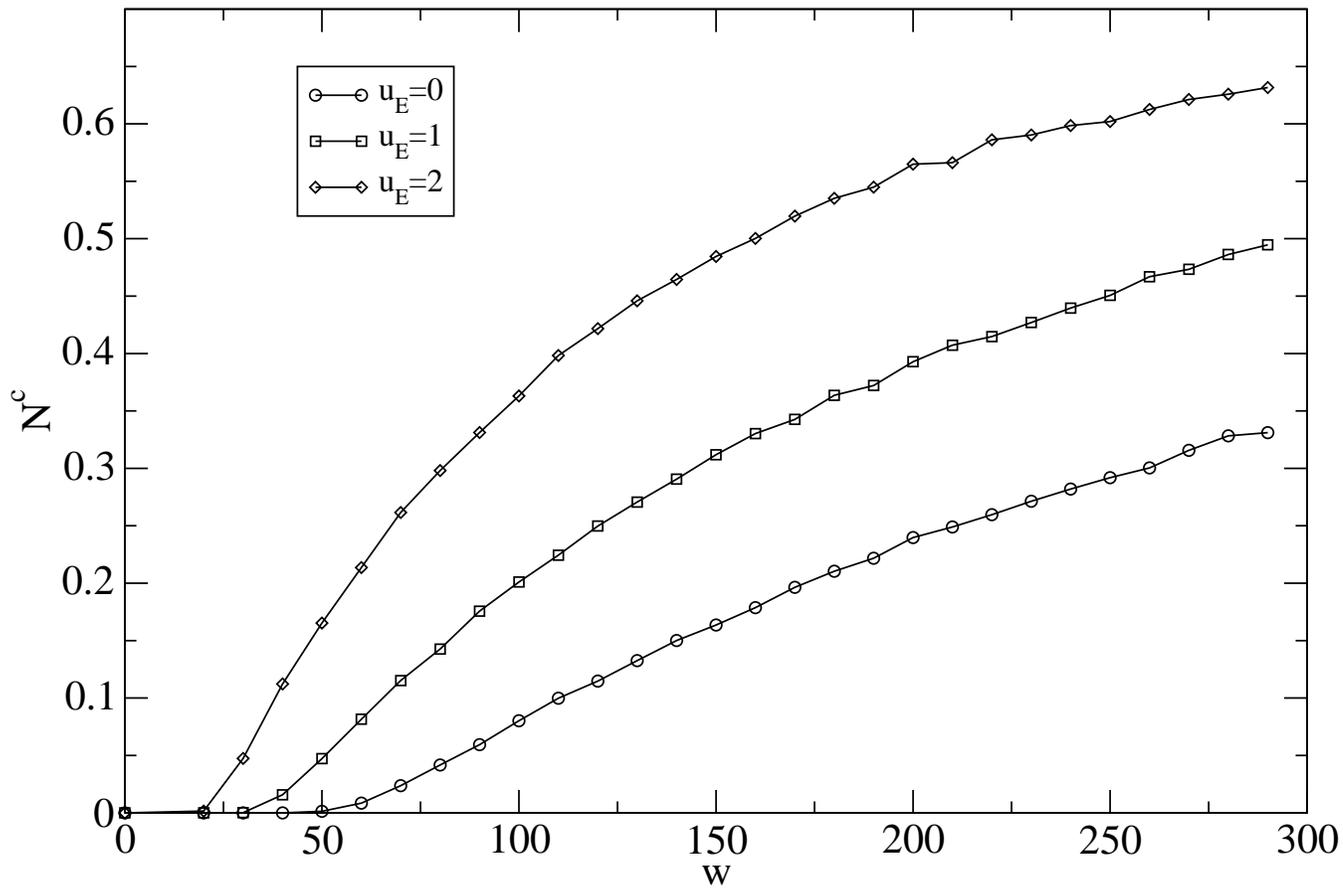}
\caption{Proportion of particles crossing the wake, $N^c$, versus
vortex strength $w$ for the non-periodic flow. The different
curves correspond to different values of the Ekman pumping, $u_E$,
as indicated in the legend. }
\label{fig:aperw}
\end{figure}

\begin{figure}
\centering{
\mbox{
\includegraphics [angle=270, width=0.5\textwidth]{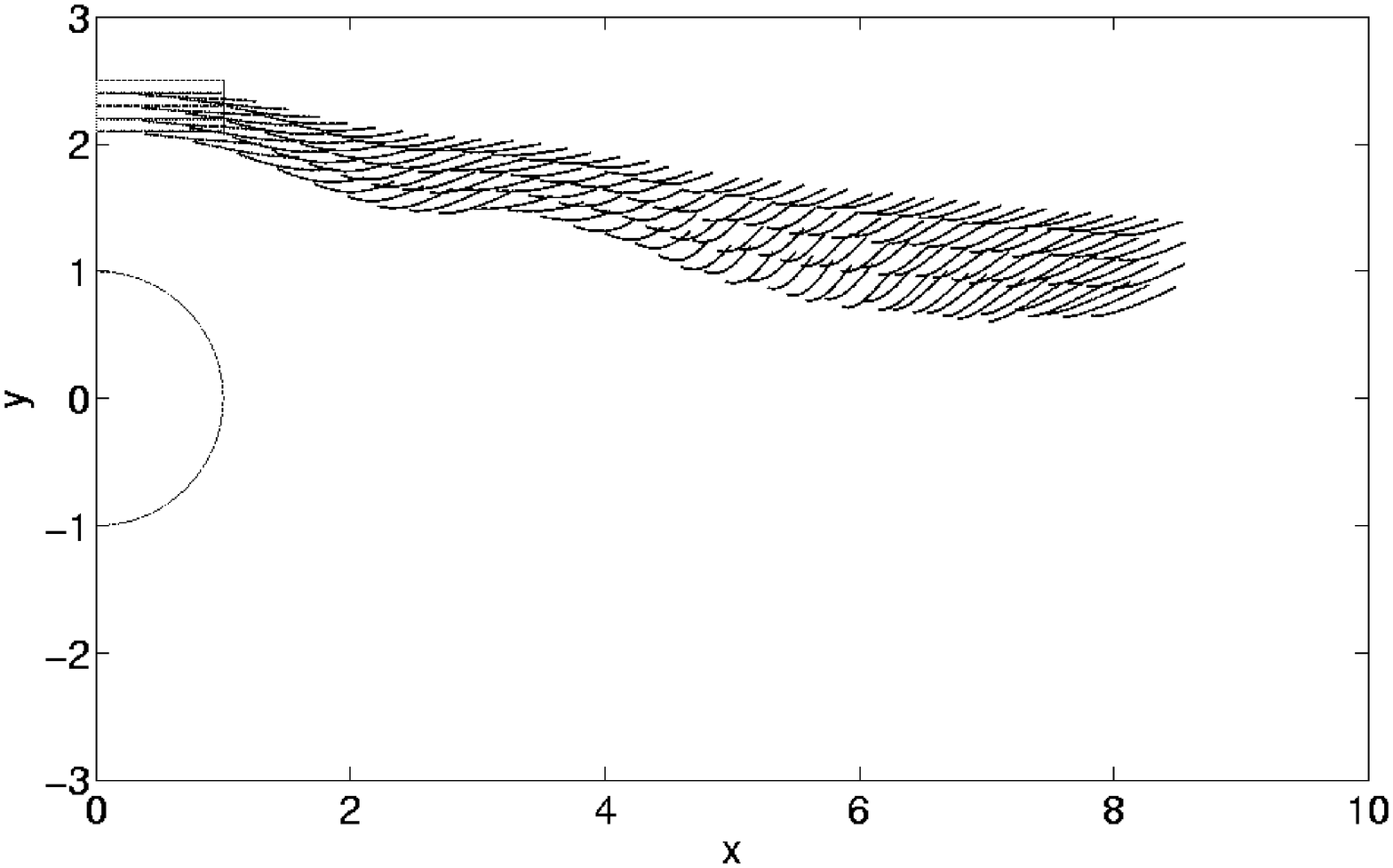}
\includegraphics [angle=270, width=0.5\textwidth]{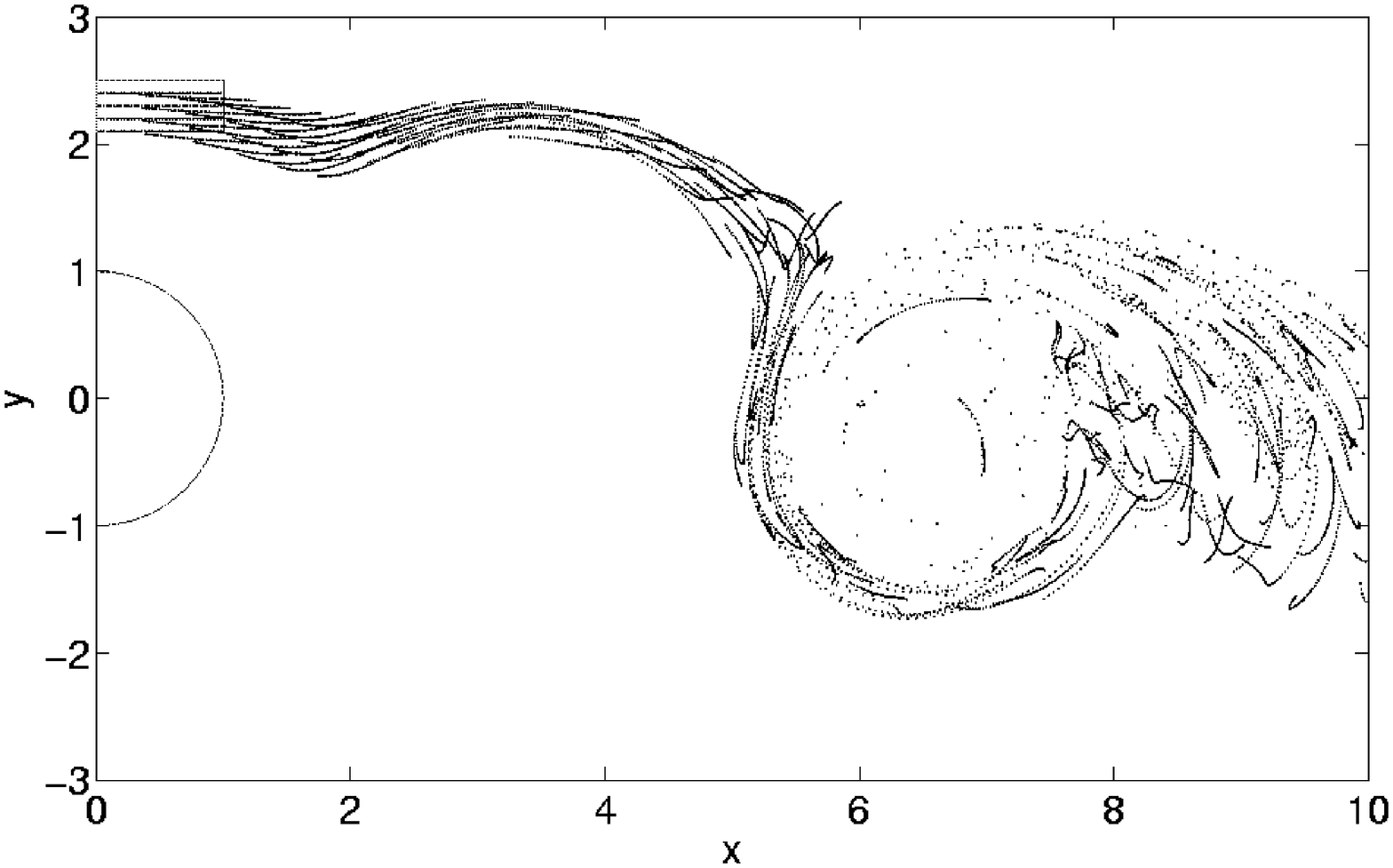}
} }\caption{Plot of the spatial distribution of the tracers in the
wake of the island for the case of the non-periodic flow and
$u_E=2$. Left: Snapshot of the distribution of the tracers at time
$0.39T_c$ for vortex strength $w=10$. Right: Snapshot of the
distribution of the tracers at time $0.39T_c$ for vortex strength
$w=200$.
\label{fig:tracersnperi}}
\end{figure}

\begin{figure}
\includegraphics [angle=0, width=1.\textwidth]{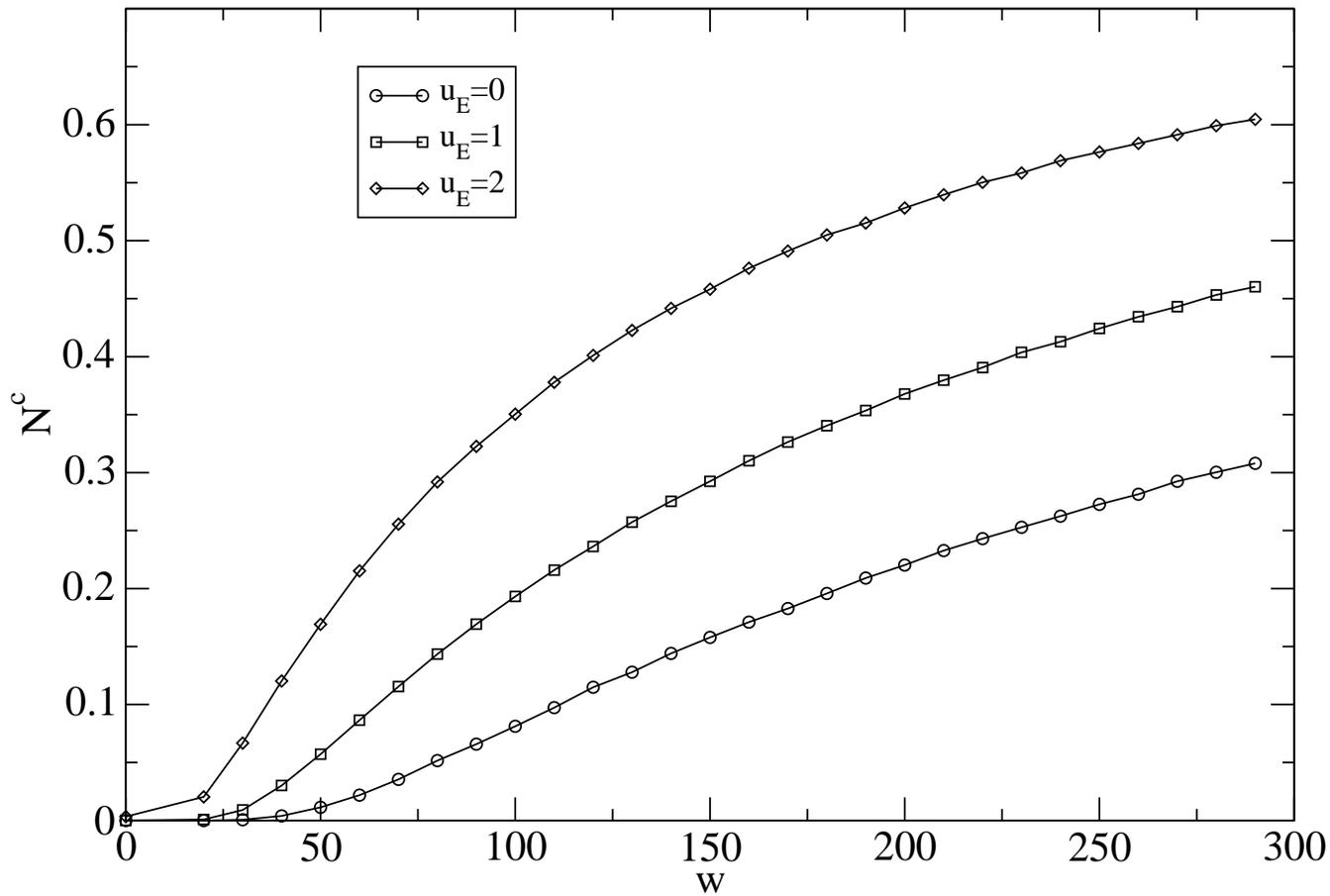}
\caption{Proportion of particles crossing the wake, $N^c$, versus
vortex strength $w$ for the periodic flow with turbulent
diffusion of the particles. The different curves correspond to
different values of the Ekman pumping, $u_E$, as indicated in the
legend. }
\label{fig:turw}
\end{figure}

\begin{figure}
\centering{
\mbox{
\includegraphics[angle=270, width=0.5\textwidth]{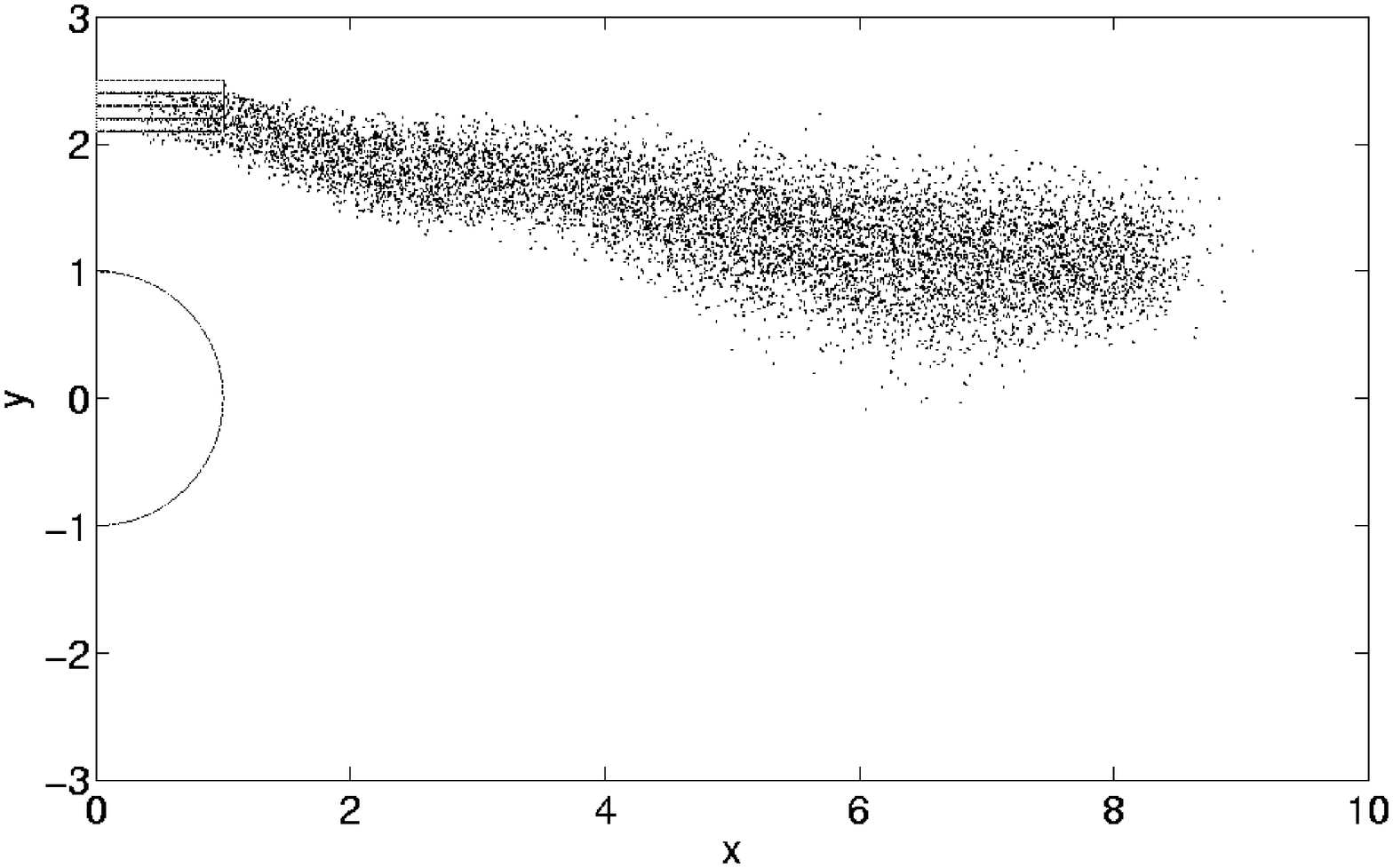}
\includegraphics[angle=270, width=0.5\textwidth]{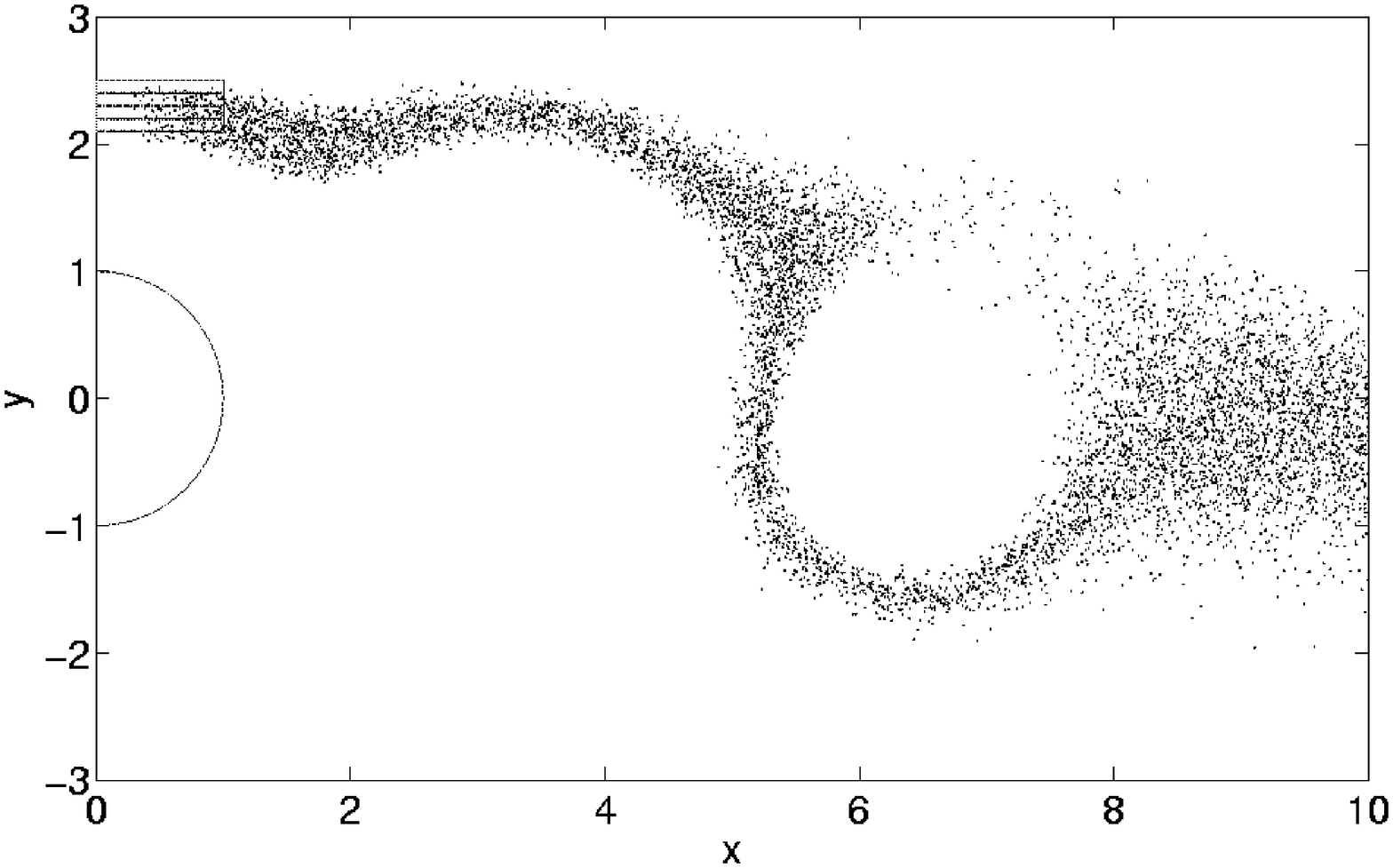}
} }\caption{Plot of the spatial distribution of tracers in the
wake of the island for the case of the periodic flow with
turbulent diffusion and $u_E=2$. Left: Snapshot of the
distribution of the tracers at time $0.39T_c$ for vortex strength
$w=10$. Right: Snapshot of the distribution of the tracers at time
$0.39T_c$ for vortex strength $w=200$.
\label{fig:tracersperiturb}}
\end{figure}

\begin{figure}
\includegraphics [angle=0, width=1.\textwidth]{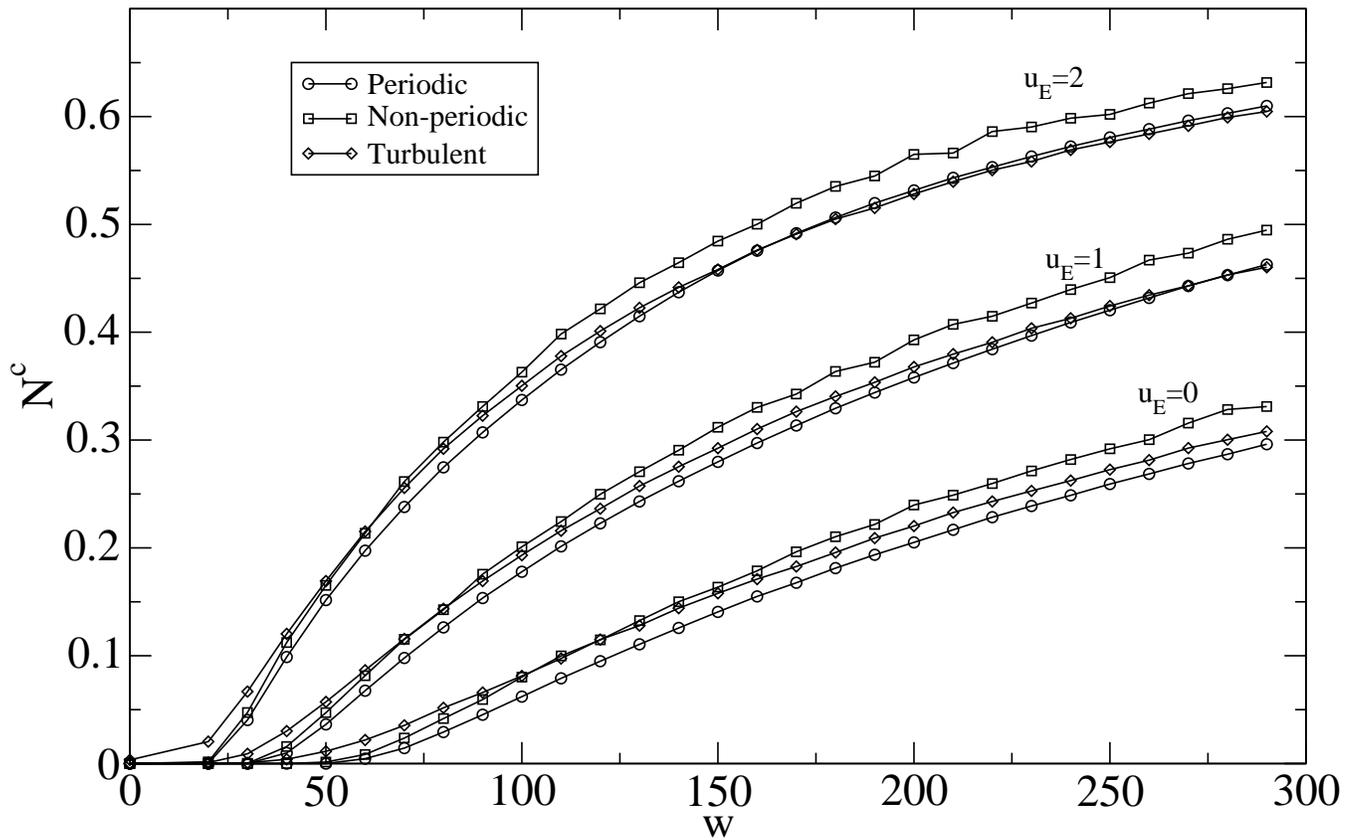}
\caption{Comparison of transport across the wake for the three
kinds of flows. The values of $u_E$ and the type of flow that
originated the data are indicated in the plot.}
\label{fig:comparacion}
\end{figure}

\begin{figure}
\includegraphics [angle=0, width=1.\textwidth]{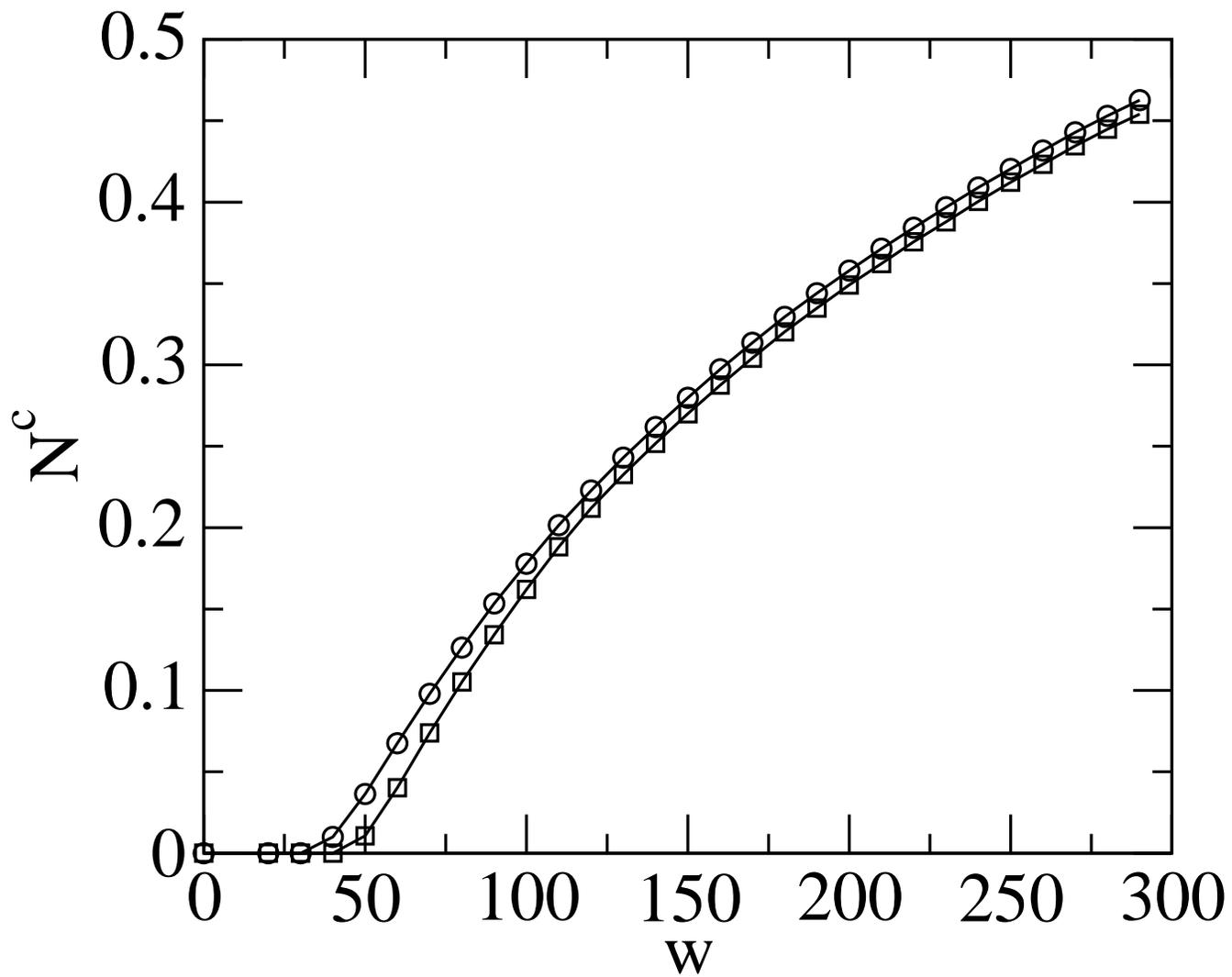}
\caption{$N^c$ versus $w$ for $u_E=1$ for the periodic flow, and
two situations: circles correspond to the proportion of particles
that cross the line $y=0$, and  squares to $y=-1$. No remarkable
differences can be observed. }
\label{fig:comparacionY}
\end{figure}

Fig. 3.  Streamlines of the flow  with Ekman flow $u_E =2$ and
other parameters, and time sequence of the snapshots, as in Figure
\ref{fig:strf}.

Fig. 4. Stable (in gray) and unstable (in black) manifolds of
the chaotic saddle in the wake of the cylinder, for the case of
the periodic flow and $u_E=2$. Left: Snapshot taken at time
$7T_c$ for vortex strength $w=10$. Right: Snapshot taken at
time $7T_c$ for vortex strength $w=200$. In the inset
we show a zoom of the manifolds in the area close to the cylinder.
 Other parameters as
described in Sect. 3. The chaotic saddle itself is closely packed
immediately behind the cylinder surface. The unstable manifold has
been plotted by releasing a large number of particles left of the
cylinder and very close to it, letting the flow to transport them
for a long time ($7T_c$ as already indicated)
 so that only the ones lasting at this time in the
wake region are still there and plotted. The stable manifold is
plotted in the same way but releasing the particles right of the
cylinder and running the flow backwards in time.

Fig. 5. The ratio of particles crossing the
wake, $N^c$, versus vortex strength $w$ in the periodic flow. The
different curves correspond to different values of the Ekman
pumping, $u_E$, as indicated in the legend.

Fig. 6. The critical values of the vortex strength, $w_c$, versus the velocity
of the Ekman pumping $u_{E}$ in the periodic flow.

Fig. 7. Plot of the spatial distribution of the tracers in the
wake of the island for the case of the periodic flow and $u_E=2$.
Left: Snapshot of the distribution of the tracers at time
$0.39T_c$ for vortex strength $w=10$. Right: Snapshot of the
distribution of the tracers at time $0.39T_c$ for vortex strength
$w=200$.

Fig. 8. Proportion of particles crossing the wake, $N^c$, versus
vortex strength $w$ for the non-periodic flow. The different
curves correspond to different values of the Ekman pumping, $u_E$,
as indicated in the legend.

Fig. 9. Plot of the spatial distribution of the tracers in the
wake of the island for the case of the non-periodic flow and
$u_E=2$. Left: Snapshot of the distribution of the tracers at time
$0.39T_c$ for vortex strength $w=10$. Right: Snapshot of the
distribution of the tracers at time $0.39T_c$ for vortex strength
$w=200$.

Fig. 10. Proportion of particles crossing the wake, $N^c$, versus
vortex strength $w$ for the periodic flow with turbulent
diffusion of the particles. The different curves correspond to
different values of the Ekman pumping, $u_E$, as indicated in the
legend.

Fig. 11. Plot of the spatial distribution of tracers in the
wake of the island for the case of the periodic flow with
turbulent diffusion and $u_E=2$. Left: Snapshot of the
distribution of the tracers at time $0.39T_c$ for vortex strength
$w=10$. Right: Snapshot of the distribution of the tracers at time
$0.39T_c$ for vortex strength $w=200$.

Fig. 12. Comparison of transport across the wake for the three
kinds of flows. The values of $u_E$ and the type of flow that
originated the data are indicated in the plot.

Fig. 13.  $N^c$ versus $w$ for $u_E=1$ for the periodic flow, and
two situations: circles correspond to the proportion of particles
that cross the line $y=0$, and  squares to $y=-1$. No remarkable
differences can be observed.

\clearpage

\clearpage

\clearpage

\clearpage

\clearpage


\clearpage

\clearpage

\clearpage

\clearpage

\clearpage

\clearpage

\clearpage

\clearpage

\clearpage

\begin{thebibliography}{}

\bibitem[\protect\citeauthoryear{Aref}{Aref}{2003}]{Aref2002}
\sur{Aref}, \fnm{H.} \date{2002}. \docu{The development of chaotic advection}.
\jou{\textit{Phys. Fluids}} \vol{\textbf{14}}, \page{1315--1325}.

\bibitem[\protect\citeauthoryear{Ar{\'\i}stegui et al.}{Ar{\'\i}stegui
et al.}{1997}]{Aristegui1997}
\sur{Ar{\'\i}stegui}, \fnm{J.},
\sur{Tett}, \fnm{P.},
\sur{Hern{\'a}ndez-Guerra}, \fnm{A.},
\sur{Basterretxea}, \fnm{M.},
\sur{Montero}, \fnm{F.},
\sur{Wild}, \fnm{K.},
\sur{Sangr{\'a}}, \fnm{P.},
\sur{Hern{\'a}ndez-Leon}, \fnm{S.},
\sur{Canton}, \fnm{M.},
\sur{Garc{\'\i}a-Braun}, \fnm{J. A.},
\sur{Pacheco}, \fnm{M.},
\mis{and} \sur{Barton}, \fnm{E. D.}
\date{1997}. \docu{The influence of island-generated eddies on chlorophyll
distribution: a study of mesoscale variation around Gran Canaria}.
\jou{\textit{Deep Sea Research I}} {\bf
44}, \page{71--96}.


\bibitem[\protect\citeauthoryear{Ar{\'\i}stegui et al.}{Ar{\'\i}stegui et
al.}{2004}]{Aristegui2004}
\sur{Ar{\'\i}stegui}, \fnm{J.},
\sur{Barton}, \fnm{E. D.},
\sur{Tett}, \fnm{P.},
\sur{Montero}, \fnm{F.},
\sur{Garc{\'\i}a-Mu{\~n}oz}, \fnm{M.},
\sur{Basterretxea}, \fnm{G.},
\sur{Cussatlegras}, \fnm{A.-S.},
\sur{Ojeda}, \fnm{A.}
\mis{and} \sur{de Armas}, \fnm{D.}
\date{2004}. \docu{Variability in plankton community
structure, metabolism, and vertical carbon fluxes along an
upwelling filament (Cape Juby, NW Africa)}. \jou{\textit{Progress in
Oceanography}} {\bf 62},
\page{95--113}.



\bibitem[\protect\citeauthoryear{Barton et al.}{Barton et
al.}{1998}]{Barton1998}
\sur{Barton}, \fnm{E. D.},
\sur{Ar{\'\i}stegui}, \fnm{J.},
\sur{Tett}, \fnm{P.},
\sur{Canton}, \fnm{M.},
\sur{Garc{\'\i}a-Braun}, \fnm{J.},
\sur{Montero}, \fnm{F.},
\sur{Nykjaer}, \fnm{L.},
\sur{Almeida}, \fnm{C.},
\sur{Almunia}, \fnm{J.},
\sur{Ballesteros}, \fnm{S.},
\sur{Basterretxea}, \fnm{G.},
\sur{Esc{\'a}nez}, \fnm{J.},
\sur{Garc{\'\i}a-Weill}, \fnm{L.},
\sur{Hern{\'a}ndez-Guerra}, \fnm{F.},
\sur{L{\'o}pez-Laatzen}, \fnm{F.},
\sur{Moliona}, \fnm{R.},
\sur{Montero}, \fnm{M. F.},
\sur{Navarro-P{\'e}rez}, \fnm{E.},
\sur{Rodr{\'\i}guez}, \fnm{J. M.},
\sur{van Lenning}, \fnm{K.},
\sur{V{\'e}lez}, \fnm{K.},
\mis{and} \sur{Wilda}, \fnm{K.}
\date{1998}. \docu{The transition zone of the Canary Current upwelling region}.
\jou{\textit{Progress in Oceanography}} {\bf
41},
\page{455--504}.



\bibitem[\protect\citeauthoryear{Barton et al.}{Barton et al.}{2004}]{Barton2004}
\sur{Barton}, \fnm{E. D.},
\sur{Ar{\'\i}stegui}, \fnm{J.},
\sur{Tett}, \fnm{P.},
\mis{and} \sur{Navarro-P{\'e}rez}, \fnm{E.}
\date{2004}. \docu{Variability in the Canary Islands area of filament-eddy exchanges}. \jou{\textit{Progress in
Oceanography}} {\bf 62}, \page{71--94}.


\bibitem[\protect\citeauthoryear{Bower et al.}{Bower et al.}{1985}]{Bower1985}
\sur{Bower}, \fnm{A. S.},
\sur{Rossby}, \fnm{H. T.}
\mis{and} \sur{Lillibridge}, \fnm{J. L.}
\date{1985}. \docu{The Gulf Stream -- barrier or blender?} . \jou{\textit{J.
Phys. Oceanogr.}} {\bf 15}, \page{26--32}.


\bibitem[\protect\citeauthoryear{Bower}{Bower}{1991}]{Bower1991}
\sur{Bower}, \fnm{A. S.}
\date{1991}. \docu{A simple kinematic mechanism for mixing fluid parcels across
a meandering jet}.
\jou{\textit{J. Phys. Oceanogr.}} \vol{\textbf{21}}, \page{173--180}.


\bibitem[\protect\citeauthoryear{Buffoni et al.}{Buffoni et al.}{1997}]{Buffoni}
\sur{Buffoni}, \fnm{G.},
\sur{Falco}, \fnm{P.},
\sur{Griffa}, \fnm{A.} \mis{and}
\sur{Zambianchi}, \fnm{E.}
\date{1997}. \docu{Dispersion processes and
residence times in a semi-enclosed basin with recirculating gyres: An
application to the Tyrhenian Sea}. \jou{\textit{J. Geophys. Res.}} {\bf 102},
\page{18699--18713}.


\bibitem[\protect\citeauthoryear{Cencini et al.}{Cencini et
al.}{1999}]{Cencini1999}
\sur{Cencini}, \fnm{M.},
\sur{Lacorata}, \fnm{G.},
\sur{Vulpiani}, \fnm{A.} \mis{and}
\sur{Zambianchi}, \fnm{E.}
\date{1999}, \docu{Mixing in a Meandering Jet: A Markovian Approximation} .
\jou{\textit{J. Phys. Oceanogr.}} {\bf 29}, \page{2578--2594}.


\bibitem[\protect\citeauthoryear{Duan and Wiggins}{Duan and Wiggins}{1997}]{Duan1997}
\sur{Duan}, \fnm{J.} \mis{and}
\sur{Wiggins}, \fnm{S.}
\date{1997}. \docu{Lagrangian transport and chaos in
the near wake of the flow around an obstacle: a numerical implementation of lobe
dynamics}. \jou{\textit{Nonlinear Processes in Geophysics}} {\bf 4},
\page{125--136}.


\bibitem[\protect\citeauthoryear{Falco et al.}{Falco et al.}{2000}]{Falco2000}
\sur{Falco}, \fnm{P.},
\sur{Griffa}, \fnm{A.},
\sur{Poulain}, \fnm{P. M.}  \mis{and}
\sur{Zambianchi}, \fnm{E.}
\date{2003}. \docu{Transport properties in the
Adriatic Sea as deduced from drifter data}. \jou{\textit{J. Phys. Oceanogr.}}
{\bf 30}, \page{2055--2071}.


\bibitem[\protect\citeauthoryear{Griffa}{Griffa}{1996}]{Griffa96}
\sur{Griffa}, \fnm{A.}
\date{1996}. \docu{Applications of stochastic particle models to oceanographic
problems}.
\mis{In}: \book{\textit{Stochastic Modelling in Physical
Oceanography}}, (\mis{eds}. \fnm{R. J.} \sur{Adler}, \fnm{P.} \sur{Muller}
\mis{and} \fnm{B. L.} \sur{Rozovskii}), \pub{Birkhauser}, \city{Boston},
\page{114--140}.


\bibitem[\protect\citeauthoryear{Jung et al.}{Jung et al.}{1993}]{Jung1993}
\sur{Jung}, \fnm{C.},
\sur{T\'el}, \fnm{T.} \mis{and}
\sur{Ziemniak}, \fnm{E.}
\date{1993}. \docu{Application of scattering chaos to
particle transport in a hydrodynamical flow}. \jou{\textit{Chaos}} {\bf
3},
\page{555--568}.


\bibitem[\protect\citeauthoryear{Mariano et al.}{Mariano et al.}{2002}]{Mariano}
\sur{Mariano}, \fnm{A. J.},
\sur{Griffa}, \fnm{A.},
\sur{{\"O}zg{\"o}kmen}, \fnm{T. M.} \mis{and}
\sur{Zambianchi}, \fnm{E.}
\date{2002}. \docu{Lagrangian analysis and predictability of coastal and ocean
dynamics}. \jou{\textit{J. Atm. Ocean. Tech.}} {\bf 19}, \page{1114--1125}.


\bibitem[\protect\citeauthoryear{Meyers}{Meyers}{1994}]{Meyers1994}
\sur{Meyers}, \fnm{S. D.}
\date{1994}. \docu{Cross-frontal mixing in a meandering jet}. \jou{\textit{J.
Phys. Oceanogr.}} \vol{\textbf{24}}, \page{1641--1646}.


\bibitem[\protect\citeauthoryear{Miller et al.}{Miller et
al.}{2002}]{Miller2002}
\sur{Miller}, \fnm{P. D.},
\sur{Pratt}, \fnm{L. J.},
\sur{Helfrich}, \fnm{K. R.},
\mis{and} \sur{Jones}, \fnm{C.K.R.T.}
\date{2002}. \docu{Chaotic transport of mass and potential vorticity for an
island recirculation}. \jou{\textit{J. Phys. Oceanogr.}} {\bf 32},
\page{80--102}.


\bibitem[\protect\citeauthoryear{Okubo}{Okubo}{1971}]{Okubo}
\sur{Okubo}, \fnm{A.}
\date{1971}. \docu{Oceanic diffusion diagrams}.
\jou{\textit{Deep-Sea Research}} \vol{\textbf{18}}, \page{789--802}.


\bibitem[\protect\citeauthoryear{Ottino}{Ottino}{1989}]{Ottino1989}
\sur{Ottino}, \fnm{J. M.}
\date{1989}. \book{\textit{The kinematics of mixing:
stretching, chaos, and transport}},
\pub{Cambridge Univ. Press}, \city{Cambridge}.


\bibitem[\protect\citeauthoryear{Pelegri et al.}{Pelegri et
al.}{2005}]{Pelegri2005}
\sur{Pelegr{\'\i}}, \fnm{J. L.},
\sur{Ar{\'\i}stegui}, \fnm{J.},
\sur{Cana}, \fnm{L.},
\sur{Gonz{\'a}lez-D{\'a}vila}, \fnm{M.},
\sur{Hern{\'a}ndez-Guerra}, \fnm{A.},
\sur{Hern{\'a}ndez-Le{\'o}n}, \fnm{A.},
\sur{Marrero-D{\'\i}az}, \fnm{M. F.},
\sur{Montero}, \fnm{M. F.},
\sur{Sangr{\`a}}, \fnm{P.},
\mis{and} \sur{Santana-Casiano}, \fnm{M.}
\date{2005}. \docu{Coupling between the open ocean and the coastal upwelling
region off northwest Africa: Water recirculation and offshore pumping of organic
matter}. \jou{\textit{J. Mar. Sys.}} {\bf 54}, \page{3--37}.


\bibitem[\protect\citeauthoryear{Rogerson et al.}{Rogerson et
al.}{1999}]{Rogerson1999}
\sur{Rogerson}, \fnm{A. M.},
\sur{Miller}, \fnm{P. D.},
\sur{Pratt}, \fnm{L. J.},
\mis{and} \sur{Jones}, \fnm{C.K.R.T.}
\date{1999}. \docu{Lagrangian motion and
fluid exchange in a barotropic meandering jet}. \jou{\textit{J. Phys.
Oceanogr.}} {\bf 29}, \page{2635--2655}.


\bibitem[\protect\citeauthoryear{Samelson}{Samelson}{1992}]{Samelson1992}
\sur{Samelson}, \fnm{R. M.}
\date{1992}. \docu{Fluid exchange across a meandering jet}. \jou{\textit{J.
Phys. Oceanogr.}} \vol{\textbf{22}}, \page{431--440}.


\bibitem[\protect\citeauthoryear{Shariff et al.}{Shariff et al.}{1992}]{Shariff1992}
\sur{Shariff}, \fnm{K. T.},
\sur{Pulliam}, \fnm{T.},
\mis{and} \sur{Ottino}, \fnm{J.}
\date{1992}. \docu{A dynamical systems analysis of kinematics in the time-periodic
wake of a circular cylinder}.
\mis{In}: \book{\textit{Vortex Dynamics and vortex methods, Proc. AMS-SIAM
Conf., Lectures in Applied Mathematics}}, (\mis{eds}. \fnm{C.}
\sur{Anderson} \mis{and} \fnm{C.} \sur{Greengard}),
\pub{American Mathematical Society}, \city{Providence}.


\bibitem[\protect\citeauthoryear{Wiggins}{Wiggins}{1999}]{Wiggins1992}
\sur{Wiggins}, \fnm{S.} \date{1992}.
\book{\textit{Chaotic transport in dynamical systems}},
\pub{Springer Verlag}, \city{New York}.


\bibitem[\protect\citeauthoryear{Ziemniak et al.}{Ziemniak et
al.}{1994}]{Ziemniak1994}
\sur{Ziemniak}, \fnm{E.},
\sur{Jung}, \fnm{C.}
\mis{and} \sur{T\'el}, \fnm{T.} \date{1994}. \docu{Tracer dynamics in open
hydrodynamical flows as chaotic scattering}. \jou{\textit{Physica  D}} {\bf 76},
\page{424}.


\end{thebibliography}
\end{document}